\documentclass[aps, preprint,superscriptaddress]{revtex4-1}
\usepackage{amsfonts}
\usepackage{amsmath}
\usepackage{amssymb}
\usepackage{graphicx}
\usepackage{times}
\usepackage{physics}
\usepackage[colorlinks=true,linkcolor=blue,citecolor=red,plainpages=false,pdfpagelabels]
{hyperref}
\usepackage{color}

\begin{document}

\title{Mesoscopic quantum superposition states of weakly-coupled 
 matter-wave solitons}

\author{Dmitriy Tsarev, Alexander Alodjants, The Vinh Ngo}
\address{National Research University for Information Technology, Mechanics and Optics (ITMO), Faculty of Laser Photonics and Optoelectronics, St. Petersburg, 197101, Russia}
\author{Ray-Kuang Lee}
\address{Physics Division, National Center for Theoretical Sciences, Hsinchu 30013, Taiwan}
\address{Institute of Photonics Technologies, National Tsing Hua University, Hsinchu 30013, Taiwan}
\address{Center for Quantum Technology, Hsinchu 30013, Taiwan}

\date{\today}

\begin{abstract}
The Josephson junctions (JJs) are at the heart of modern quantum technologies and metrology. In this work we establish quantum features of an atomic soliton Josephson junction (SJJ) device, which consists of two weakly-coupled condensates with negative scattering length. The condensates are trapped in a double-well potential and elongated in one dimension. Starting with classical field theory we map for the first time a two-soliton problem onto the effective two-mode Hamiltonian and perform a second quantization procedure. Compared to the conventional Bosonic Josephson junction (BJJ) condensate system, we show that the SJJ-model in quantum domain exhibits unusual features due to its effective nonlinear strength proportional to the square of total particle number, $N^2$. A novel self-tuning effect for the effective tunneling parameter is also demonstrated in the SJJ-model, which depends on the particle number and rapidly vanishes as the JJ population imbalance increases. 
The formation of entangled Fock state superposition is predicted for the quantum SJJ-model, revealing dominant $N00N$-state components at the ``edges" for $n=0, N$ particle number. We have shown that the obtained quantum state is more resistant to few particle losses from the condensates if tiny components of entangled Fock states are present in the vicinity of the major $N00N$-state component.
This peculiarity of  the  quantum  SJJ-model  establishes an important difference from its semiclassical analogue obtained in the framework of Hartree approach.
Our results are confirmed by studying the first and $N$-order Hillery-Zubairy criteria applied for studying multiparticle entanglement and planar spin squeezing. The Einstein-Podolsky-Rosen (EPR) quantum steering represents an important prerequisite for the crossover to the mesoscopic superposition Schr{\"o}dinger-cat and/or $N00N$-states. The feasibility in observation for these predicted states of the SJJ-model in the experiments is also discussed by taking into account one- and three- body losses for lithium condensates.
\end{abstract}


\maketitle
\newpage

\section{Introduction}

The Josephson junctions (JJs) represent an indispensable tool for current quantum technologies. The Josephson effect, that is at the core of various JJ-devices, presumes the macroscopic quantum effect of particle tunneling between two (or more) junctions; it was initially discovered in superconductors with Cooper pairs~\cite{Josephson1974}. The JJs, which exhibit superfluidity, were demonstrated with $\mathrm{{}^3He}$, $\mathrm{{}^4He}$ environments~\cite{Pereverzev1997,Davis2002,Sukhatme2001}. The atomic, so-called Bosonic Josephson junctions (BJJs), were shown for weakly-coupled atomic Bose-Einstein condensates (BECs)~\cite{Leggett2001,Albiez2005}.

Since experimentally accessible temperatures of atomic BECs are very low (tens of microKelvins and below), the dynamical regimes reveal an interplay between coherent Josephson tunneling and nonlinearity characterising weak atom-atom interaction in Born approximation~\cite{Milburn1997,Raghavan1999}. The exciton-polariton BECs in narrow band semiconductor microstructures allow to obtain JJs at much higher temperatures like a few of Kelvins~\cite{Abbarchi2013,Lagoudakis2010}. The photons possess tunneling between the contacts; however, in this case dissipation and temperature dependent phenomena must be taken into account~\cite{Lebedev2017}. Strictly speaking, such tunnelly-coupled waveguides represent purely photonic analogues of JJs far from thermal equilibrium~\cite{Maier1995,Leksin2003}. 

On the contrary, the low temperatures promote the observation of purely quantum phenomena with JJs, such as macroscopic or mesoscopic superposition Schr{\"o}dinger-cat (SC) states~\cite{Cirac1998,Huang2006,Olsen2010,Reid2011,Salasnich2011}, entanglement, and spin squeezing in ultracold atomic ensembles~\cite{Reid2011,Sorensen2001,Sorensen2001a,Reid2012,Vitagliano2018,Reid2018, Maa2011}. It is also important to mention so-called steerable states, which represent a physically specific and mathematically strict subset of the set of entangled (non-separable) states~\cite{Jones2007}. In particular, such states mean the ability of Alice to remotely prepare Bob's state. As shown in~\cite{Reid2012}, EPR (Einstein-Podolsky-Rosen) steering for BJJs appears in atomic condensates with negative atom-atom scattering length.

From fundamental point of view, BJJ-devices may form macroscopic SC-states~\cite{Leggett1984}. Some recent results associated with symmetry (rotational and inversion) are obtained for superposition states in~\cite{Castanos2012, Castanos2020}. Practically, BJJ quantum features are promising for realization of macroscopic quantum computation~\cite{Byrnes2012,Byrnes2011}, as well as for quantum metrology and sensing purposes~\cite{Pezze2018}. In particular, planar quantum squeezing, discussed in the framework of atomic JJs, allows to enhance the metrological sensitivity of measurements for arbitrary phase shifts~\cite{Puentes2013}. Noteworthy, SC-states and especially maximally path-entangled $N$-particle $N00N$-states represent a paramount tool to achieve the phase measurement at the Heisenberg sensitivity level~\cite{Dowling2008}.

The experimental generation of $N00N$-states with a large number of particles, $N$, is one of the important and non-trivial problems, which significantly affects the state of art of current quantum metrology~\cite{Afek2010,Chen2010,Merkel2010}. Investigating various aspects of appropriate $N00N$-states is imperative in the presence of decoherence and losses, which can rapidly destroy $N00N$-states~\cite{Dorner2009, Vogel2015}. For example, in the presence of losses balanced $N00N$-states are non-optimal for the implementation in quantum metrology~\cite{Huver2008, Dobrzanski2009}. Designing macroscopic quantum states, which may be robust to losses, represents an essential and challenging task for current quantum technologies with atomic systems, cf.~\cite{Sinatra2017}.

In this work we propose to create macroscopic entangled (Fock) states by means of tunnely-coupled bright atomic solitons, which demonstrate some useful features for quantum metrology purposes, cf.~\cite{Tsarev2018, Tsarev2019, Raghavan2000}. In particular, as reported in~\cite{Tsarev2019}, quantum atomic solitons allow to surpass the Heisenberg limit and obtain $N^{-3/2}$ accuracy in nonlinear phase shift estimation even with coherent probes in the framework of nonlinear quantum metrology. To overcome this limit we suggested SC- and $N00N$-states formation in the scheme with two weakly-coupled condensate bright solitons. 

However, in~\cite{Tsarev2018, Tsarev2019} we restricted ourselves by the Hartree approximation that cannot fully provide the advantages of the solitonic model in a quantum domain, cf.~\cite{Raghavan2000}. In this work we show that the system of coupled atomic solitons represents a new quantum soliton Josephson junction (SJJ) device with a nontrivial physical behavior in comparison with a conventional BJJ-device based on atomic condensates, which possess Gaussian-shape wave functions. In particular, paper aims to demonstrate that SJJs are suitable for the generation of entangled superposition Fock states, which may be resistant to some weak losses.

Remarkably, quantum solitons represent the Lieb-Liniger model, which was discussed in the literature for a long time ago, see e.g.~\cite{Lieb1963,Haus1989,Zill2015}. Various aspects of solitons quantum fluctuations related to quadrature squeezing, entanglement, and photon number statistics were studied in a quantum optics domain~\cite{Drummond1987,Friberg1996,Spalter1998,Lai2009}. However, optical solitons represent systems containing a large number of particles far from thermal equilibrium. In this sense, atomic solitons favorably differ from optical analogues and could be recognized as the best candidates for the experimental realization of macroscopic quantum superposition states. 

The progress in theory and experiment achieved with matter-wave bright solitons is discussed and summarized in~\cite{Kevrekidis2008}. The bright solitons were experimentally demonstrated in several labs with atomic BECs possessing negative scattering length and allowing a moderate number of particles (from several tens to thousands)~\cite{Strecker2002,Khaykovich2002}. The collisions for such solitons were recently shown in~\cite{Nguyen2014}. Notably, the observation of bright matter-wave solitons, which admit collapsing phenomena due to attractive inter-particle interaction, represents a non-trivial task for experimenters, see e.g. Lev Khaykovich's article in Ref.~\cite{Kevrekidis2008}. Alternatively, bight solitons may be formed in periodic structures by manipulating the sign of effective particle mass~\cite{Eiermann2004}. However, exploiting a macroscopically large particle number (thousands and more) in solitons leads to various (many-body) losses~\cite{Weiss2016}. In this sense, solitons with mesoscopic number of particles (few hundreds and less) represent a primary interest. Some of recent proposals with quantum atomic solitons are established in~\cite{Carr2004,Malomed2006,Gardiner2012}.

Finally, let us mention mesoscopic atomic Bose-systems, so-called quantum droplets, which occur by means of manipulation with scattering atomic length, and posess intriguing quantum features~\cite{Cabrera2018}. Ferrier-Barbut et al. in~\cite{Pfau2016} demonstrated the important influence of quantum fluctuations on droplet stabilization beyond the mean-field level that occur for strong dipolar Bose gases containing several hundreds of particles. Physically, beyond-mean-field energy correction in low dimensions plays an essential role for quantum droplet formation and behavior~\cite{Petrov2016}. In particular, as predicted in~\cite{Liu2019, Tylutki2020}, quantum droplets represent soliton-like (bell-shape) objects and consist of two one-dimensional Bose-Bose mixtures with competing (repulsive and attractive) interactions.

In this work we establish a new approach to the two-soliton (SJJ) problem mapping it onto the effective two-mode Hamiltonian and performing a second quantization procedure.

The paper is arranged as follows. In Sec.~2 we discuss a classical field approach to coupled BECs possessing negative scattering length. We represent a general approach for tunneling BECs in elongated traps in Sec.~2.1. Important peculiarities and limitations for the JJ-models are also discussed. In Sec.~2.2 we represent a familiar BJJ-model problem valid for atomic condensates with a Gaussian wave function. We implement this model through the paper for validation of the results obtained with the SJJ-model. Then in Sec.~2.3 we establish the SJJ-model that operates with coupled 1D bright solitons. The Hamilton formalism is used to find mean-field equations. The important features of these equations are revealed in a steady state. The quantum approach to the BJJ- and SJJ-models is given in Section 3. The macroscopic properties of the energy spectrum for superposition states are analyzed in the framework of the Hartree approximation. In Section 4 the full quantum theory of the SJJ-model is presented. The peculiarities of the quantum energy spectrum are discussed in detail. We elucidate the quantum features of entanglement and EPR steering in the crossover region where the Hartree approach is violated. The experimental conditions for quantum superposition state formation in the presence of many-body losses are discussed in Sec.~4.3. We also analyze the influence of few particle losses on predicted condensate states in the framework of the fictitious beam splitter approach, cf.~\cite{Rubin2007}. Finally, in Conclusion we summarize the results obtained.

\section{Classical field models for coupled atomic condensates}

\subsection{General approach to JJ-models}

The atomic JJ system can be created basically by means of appropriate condensate trapping. In various experiments with JJ the BEC is confined in a potential $V=V_H+V_W$, where $V_H$ is familiar 3D harmonic trapping potential, $V_W$ is responsible for condensate double-well confinement~\cite{Albiez2005,Gati2007}. The aim of such a trapping is to split the condensate into two parts which possess tunneling through the barrier created by $V_W$ potential.

\begin{figure}[t]
\center{\includegraphics[width=0.5\linewidth]{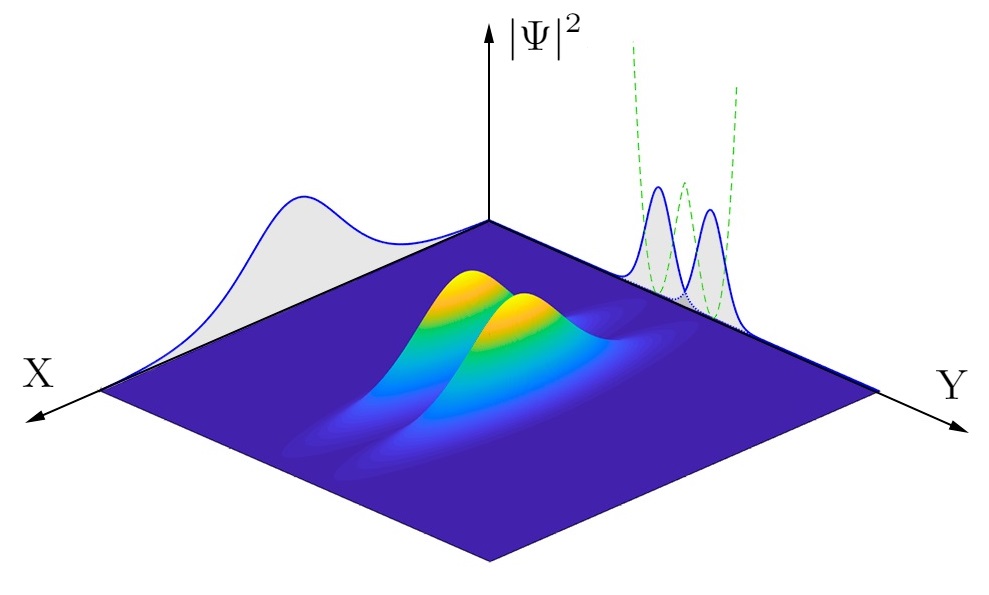}}
\caption{Sketch of probability density distribution $|\Psi|^2$ vs. spatial coordinates $X$ and $Y$ for the coupled condensates trapped in a double-well potential (dashed green curve). Shadow regions display BEC projections in $XY$- and $YZ$-planes, which demonstrate Gaussian distribution along $Y$-axis. In $X$-dimension the distribution represents secant and Gaussian for the SJJ- and BJJ-models, respectively.}
\label{FIG:W-pot}
\end{figure}

The scheme of possible realization of JJ-device with two weakly coupled condensates is evident from Fig.~\ref{FIG:W-pot}, where for clarity we depicted the probability distribution for 2D condensates. In particular, the JJ platform consists of two highly asymmetric (cigar-shaped) condensates elongated in $X$ direction and trapped in a double-well potential (green line in Fig.~\ref{FIG:W-pot}) allowing some distance $d$ between trap centers. In experiments with dipole traps $d$ is few micrometers~\cite{Gati2007}. The BEC particles possess an interaction that we suppose attractive in this work. This assumption allows to examine Gaussian (BJJ-model) and bright soliton (SJJ-model) solutions for the condensate wave functions in the framework of the same physical set-up illustrated in Fig.~\ref{FIG:W-pot}.
 
We represent the condensate wave function $\Psi(r_{\perp},x,t)$ relevant to the scheme in Fig.~\ref{FIG:W-pot} in a superposition state as 
\begin{equation}\label{CondWF}
\Psi(r_{\perp},x,t)=\Phi_1(r_{\perp}-d/2)\Psi_1(x,t) +\Phi_2(r_{\perp}+d/2)\Psi_2(x,t), 
\end{equation}
where $\Phi_{1,2}(r_{\perp}+ d/2)$ and $\Phi_{2}(r_{\perp} - d/2)$ are spatial (time independent) wave functions of condensates in transverse plane on either side of the barrier.

In Eq.~\eqref{CondWF} $\Psi_1\equiv\Psi_1(x,t)$ and $\Psi_2\equiv\Psi_2(x,t)$ are two condensate wave functions which characterize dynamical properties of JJ system and obey coupled Gross-Pitaevskii equations (cf.~\cite{Raghavan2000})
\begin{subequations}\label{Lagr}
\begin{eqnarray}
i\frac{\partial}{\partial t}\Psi_1&=&-\frac{1}{2}\frac{\partial^2}{\partial x^2}\Psi_1 - u\left|\Psi_1\right|^2\Psi_1 + \frac{1}{2}\nu^2x^2\Psi_1 - \kappa\Psi_2;\\
i\frac{\partial}{\partial t}\Psi_2&=&-\frac{1}{2}\frac{\partial^2}{\partial x^2}\Psi_2 - u\left|\Psi_2\right|^2\Psi_2 + \frac{1}{2}\nu^2x^2\Psi_2 - \kappa\Psi_1,
\end{eqnarray}
\end{subequations}
where $u=2\pi|a_{sc}|/a_{\perp}$ characterizes Kerr-like (focusing) nonlinearity, $a_{sc}<0$ is the s-wave scattering length for attractive particles, $a_{\perp}=\sqrt{\hbar/m\omega_{\perp}}$ is a characteristic trap scale, and $m$ is particle mass.

In Eq.~\eqref{Lagr} $\kappa\equiv{\left|\cal K\right|}/\omega_{\perp}>0$ denotes an absolute value of the tunneling rate (${\cal K}<0$), normalised on radial frequency, $\omega_{\perp}$; ${\cal K}$ depends on the condensate wave functions $\Phi_{1,2}(r_{\perp}\pm d/2)$ overlapping in YZ-plane; $\nu=\omega_x/\omega_{\perp}$ is a trap asymmetry parameter, $\omega_x$ and $\omega_{\perp}$ are harmonic trap frequencies in $X$-axis and radial direction, respectively.

In the framework of field theory it is possible to consider a Hamilton function corresponding to Eqs.~\eqref{Lagr}, which looks like (cf.~\cite{Tsarev2018,Tsarev2019})
\begin{equation}\label{Ham_cl}
H=\sum_{j=1}^2 \left(\frac{1}{2}\left|\frac{\partial \Psi_j}{\partial x}\right|^2-\frac{u}{2}\left|\Psi_j\right|^4+
\left|\Psi_j\right|^2 V_{H}(x)\right) - \kappa\left(\Psi_1^*\Psi_2+\Psi_1\Psi_2^*\right),
\end{equation}
where $V_{H}(x)=\frac{1}{2} \nu^2 x^2$ is trapping potential in $X$ direction. In~\eqref{Lagr},~\eqref{Ham_cl}, $\Psi_1$ and $\Psi_2$ obey a normalization condition
\begin{equation}\label{Norm}
\int_{-\infty}^\infty(|\Psi_1|^2+|\Psi_2|^2)dx=N, 
\end{equation}
where $N$ is the average total number of particles. 
 
In Eqs.~\eqref{Lagr}-\eqref{Norm} we also propose rescaled (dimension-less) spatial and time variables, which are: $x,y,z \rightarrow x/a_{\perp},y/a_{\perp},z/a_{\perp}$ and $t \rightarrow \omega_{\perp} t$, cf.~\cite{Tsarev2018,Tsarev2019,Raghavan2000}. In other words, all lengths and time variables in the work are given in $a_{\perp}$ and $\omega_{\perp}^{-1}$ units respectively.

Let us briefly discuss limitations for Eqs.~\eqref{CondWF}-\eqref{Ham_cl} and conditions for bright atomic soliton formation that we are interested in this work. In particular, if the number of condensate particles, $N$, is not too large, spatial wave functions $\Phi_{1,2}(r_{\perp}\pm d/2)$ may be considered as Gaussian and still unchanging within appropriate time scales. Typically, this assumption is valid for weakly-coupled condensates possessing a moderate number of particles, $N$. In other words, Eq.~\eqref{CondWF} represents some ansatz for the wave function that provides so-called two-mode approximation used in this work. In particular, as it is shown in~\cite{Anglin2001}, characteristic trap scale, $a_{\perp}$, should be larger than nonlinear strength, $N|a_{sc}|$, in two-mode approximation. The results obtained in the experiments with condensate BJJs manifested validity of the two-mode approach up to few thousand particles~\cite{Albiez2005,Raghavan1999,Gati2007}.

However, condensates with attractive interaction between the particles possess instability and collapse if the number of atoms exceeds some critical value, $N_c$; Refs.~\cite{Elef2000, Eigen2016} exhibit recent experimental efforts in this field. The critical particle number is $N_c=0.67 a_{\perp}/|a_{sc}|$,~\cite{Kevrekidis2008}. In particular, for condensate solitons with $\mathrm{{}^7Li}$ atoms possessing the negative scattering length, the effective nonlinear parameter is $uN_c\approx4.2$. In early experiment~\cite{Khaykovich2002} authors demonstrated that $N$ is limited by the the number of atoms $5.2\times 10^3$ in the soliton providing $N_c|a_{sc}|\simeq1.105 \mu m$. At the same time, it is shown that there exists some parameters window, where 1D bright soliton avoids collapsing. 

Another way to obtain bright atomic solitons is based on formation of band gap solitons occurring due to the condensate confinement in a periodical optical lattice. Although the condensate possesses the positive scattering length, the negative effective mass $m_{eff}<0$ of the particles that forms at the edge of Brillouin zone provides bright soliton occurrence in this case~\cite{Pearl2003}. In~\cite{Eiermann2004} authors demonstrated band gap bright soliton formation with $N\simeq250$ rubidium atoms interacting repulsively. Thus, the number of particles, $N$, in the soliton can be estimated from 
\begin{equation}\label{Sol}
N= \frac{a_{\perp}^2m}{1.5 |m_{eff}|x_0a_{sc}},
\end{equation}
where $x_0$ is the soliton width~\cite{Eiermann2004}. From~\eqref{Sol} follows that the number of particles, $N$, may be reduced if we can enhance the value of scattering length $a_{sc}$ by means of Feshbach resonance method, cf.~\cite{Cheng2010}, or with tailoring the effective atom mass, $m/|m_{eff}|$, in the lattice, cf.~\cite{Sedov2014}. The ratio $m/|m_{eff}|$ was equal to $10$ in~\cite{Eiermann2004}. To be more specific, below we restrict ourselves by the matter-wave bright solitons with the negative scattering length.

\subsection{Classical BJJ-model}

Let us assume that the wave function shape in $X$-dimension is Gaussian and looks like 
\begin{equation}\label{anzGauss}
\Psi_j=\frac{\nu^{1/4}\sqrt{N_j}}{\pi^{1/4}}e^{-\nu\,x^2/2}e^{i\theta_j},
\end{equation}
where $N_j$ and $\theta_j$ are the average particle number and phase of $j$-th condensate, respectively. The approach discussed here is valid at zero temperature in thermodynamic limit when the number of particles and occupied volume are extremely (infinitely) large but their ratio is finite. The losses and non-equilibrium phenomena are neglected.

Substituting~\eqref{anzGauss} into~\eqref{Ham_cl} and omitting constant energy terms proportional to $N$ for effective classical Hamiltonian $H_{BJJ}=\int_{-\infty}^\infty H\,dx$ one can obtain \begin{equation}\label{H_eff_G}
H_{BJJ}=-\frac{u\sqrt{\nu}}{2\sqrt{2\pi}}\left(N_1^2+N_2^2\right) - 2\kappa\sqrt{N_1N_2}\cos[\theta].
\end{equation}
It is instructive to represent the Hamiltonian~\eqref{H_eff_G} in terms of new variables $z=(N_2-N_1)/N$ and $\theta=\theta_2-\theta_1$. Omitting the constant energy term for~\eqref{H_eff_G} we obtain 
\begin{equation}\label{H_BJJ}
H_{BJJ}=\kappa N\left(-\frac{\lambda}{2}z^2 - \sqrt{1-z^2}\cos[\theta]\right),
\end{equation}
where $z$ and $\theta$ are normalized population imbalance and phase difference, respectively. In Eq.~\eqref{H_BJJ} we introduce the effective nonlinear parameter 
\begin{equation}\label{lambda}
\lambda=\frac{\sqrt{\nu}\,uN}{2\sqrt{2\pi}\,\kappa}\equiv\frac{\sqrt{\nu}\,uN\, \omega_{\perp}}{2\sqrt{2\pi}\, \left|{\cal K}\right|}. 
\end{equation}

The Eq.~\eqref{H_BJJ} with the key parameter $\lambda$ represents a conventional JJ Hamiltonian and characterizes an atomic condensate BJJ-model intensively studied in the framework of various quantum and nonlinear features occurring as a result of tunneling and inter-particle interaction interplay~\cite{Raghavan1999,Cirac1998,Olsen2010,Reid2011,Skott1992}.

It is worth noticing that the Hamiltonian~\eqref{H_BJJ} is invariant under the transformation $\kappa\rightarrow -\kappa$ and $\theta\rightarrow \theta+\pi$. However, as it is argued in [70], physically correct tunneling rate $\cal K$ must be negative, cf.~\cite{Tsarev2018, Tsarev2019, Raghavan2000}.

Now let us consider the case when two condensates in $X$-dimension may be described by bright soliton solutions. We suppose that condensate trapping in $X$-dimension is weak enough that can be realized for highly elongated condensates obtained by means of asymmetric trapping potential with $\nu<<1$, see Fig.~\ref{FIG:W-pot}. Thereafter, we completely neglect the trapping term in Eq.~\eqref{Ham_cl}, cf.~\cite{Raghavan2000}.

If coupling parameter $\kappa$ is switched-off, 
the two condensates behave independently possessing non-moving bright soliton solutions in $X$-dimension, which are 
\begin{equation}\label{anz}
\Psi_j=\frac{N_j\sqrt{u}}{2} sech\left[\frac{uN_j}{2}x\right]e^{i\theta_j}.
\end{equation}
Substituting~\eqref{anz} into~\eqref{Ham_cl} for effective classical Hamiltonian $H_{SJJ}=\int_{-\infty}^\infty Hdx$ one can obtain 
\begin{equation}\label{H_eff_0}
H_{SJJ}=-\frac{u^2}{24}\left(N_1^3+N_2^3\right) - \frac{4\kappa N_1N_2}{N}I(z)\cos[\theta],
\end{equation}
where 
\begin{equation}\label{I}
I(z)=\int_{0}^\infty\frac{dx}{\cosh^2[x]+\sinh^2[zx]}\approx 1-0.21z^2.
\end{equation}

The Hamiltonian~\eqref{H_eff_0} in terms of $z$ and $\theta$ variables reads as 
\begin{equation}\label{H_eff}
H_{SJJ}=\kappa_{eff} N\left(-\frac{\Lambda_{eff}}{2}z^2 - \sqrt{1-z^2}\cos[\theta]\right),
\end{equation}
and looks similar to Eq.~\eqref{H_BJJ} obtained for the BJJ-model. In~\eqref{H_eff} we introduced the effective nonlinear parameter 
\begin{subequations}\label{Lambda}
\begin{equation}
\Lambda_{eff}=\frac{u^2N^2}{16\kappa_{eff}}\equiv\frac{\Lambda}{(1-0.21z^2)\sqrt{1-z^2}},
\end{equation}
where
\begin{equation}
\Lambda\equiv\frac{u^2N^2}{16\kappa}=\frac{u^2N^2\,\omega_{\perp}}{16 \left|{\cal K}\right|} 
\end{equation}
\end{subequations}
 is the vital parameter that determines various regimes of soliton dynamics. 

Eq.~\eqref{H_eff} derived for the SJJ-model can be understood in the framework of the BJJ-model Hamiltonian~\eqref{H_BJJ} with tunneling rate, $\kappa_{eff}\equiv \kappa (1-0.21z^2)\sqrt{1-z^2}$, that depends on solitons population imbalance, $z$. The nonlinear effects in the tunneling process vanish for $z^2<<1$. In this limit the BJJ- and SJJ-models possess the same features.

However, when $z^2\rightarrow 1$, the effective tunneling rate, $\kappa_{eff}$, goes to zero and effective parameter $\Lambda_{eff}$ essentially increases. In other words, we deal here with \textit{self-tuning effect} for effective tunneling parameter, $\kappa_{eff}$, that establishes various regimes for interacting solitons. As we can see below, this limit is of particular interest for the quantum (superposition) states formation, which we specify in the paper.

The equations $\dot{\theta}=-\partial H_{SJJ}/\partial z$ and $\dot{z}=\partial H_{SJJ}/\partial \theta$ for canonical variables, $z$ and $\theta$, can be obtained with the Hamiltonian~\eqref{H_eff} as~\cite{Tsarev2018,Raghavan2000}:
\begin{subequations}\label{eq_class}
\begin{eqnarray}
\dot{z}&=&(1-z^2)(1-0.21z^2)\sin[\theta];\\
\dot{\theta}&=&\Lambda z - 2z(1.21-0.42z^2)\cos[\theta].
\end{eqnarray}
\end{subequations}

In Eq.~\eqref{eq_class} dots denote derivatives with respect to effective (dimensionless) time $\tau=\kappa N t$. These equations permit non-trivial steady-state solutions 
\begin{subequations}\label{SC}
\begin{eqnarray}
z_{\pm}&=&\pm \sqrt{\frac{1}{0.42}(1.21-\frac{\Lambda}{2})};\label{SC_a}\\
\cos[\theta]&=&1\label{SC_b},
\end{eqnarray}
\end{subequations}
which are valid for $1.58<\Lambda\leq2.42$. For another specific value of phase $\theta$ we have 
\begin{subequations}\label{N00N}
\begin{eqnarray}
z_{\pm}&=&\pm1;\\
\cos[\theta]&=&\frac{\Lambda}{1.58}\label{N00N_b},
\end{eqnarray}
\end{subequations}
that requires $0<\Lambda\leq 1.58$. 

It is important that classical mean-field theory admits realization of one of solutions: either with $z_+$ or with $z_-$. On the contrary, the quantum approach, that we are going to discuss below, permits the simultaneous realization of these states, being a clearly superposition state.

\section{Quantum approach to JJ-models}

\subsection{Effective Hamiltonian quantization} 

The quantum theory approach proposes a second quantization procedure for the classical effective Hamiltonians~\eqref{H_BJJ},~\eqref{H_eff}, and relevant field amplitudes. Notice that especially for nonlinear Hamiltonians containing phase-dependent terms this procedure can be non-unique because of the absence of rigorous phase quantization procedure and non-commutativity of relevant operators. However, as known, various approaches to Hamiltonian quantization permit difference in terms by the factor proportional to $1/N$ and less, which is negligibly small for large $N$. This difference occurs due to non-commutativity of annihilation and creation operators forming the Hamiltonian.

To be more specific, we start with quantization of the classical Hamiltonian~\eqref{H_BJJ} for the BJJ-model. The quantization procedure that we propose here is based on the approach described in~\cite{Leggett2011}. We construct a quantum Hamiltonian from Eq.~\eqref{H_BJJ} by introducing particle number operators $\hat{N}_1=\hat{a}^\dag\hat{a}$ and $\hat{N}_2=\hat{b}^\dag\hat{b}$, respectively; one can define the reduced particle number difference operator $\hat{z}=\frac{1}{N}\left(\hat{b}^\dag\hat{b}-\hat{a}^\dag\hat{a}\right)$, where $\hat{a}$ ($\hat{a}^\dag$) and $\hat{b}$ ($\hat{b}^\dag$) are usual Bosonic annihilation (creation) operators for condensate modes in two-mode approximation, cf.~\cite{Leggett2011}. Then one can suppose in~\eqref{H_BJJ} that $\sqrt{1-z^2}\cos[\theta]$ corresponds to $\frac{1}{N}\left(\hat{a}^\dag\hat{b}+\hat{b}^\dag\hat{a}\right)$ if we assume decomposition $\hat{a}=\sqrt{\frac{N}{2}}\sqrt{1-\hat{z}}e^{i\hat{\theta}/2}$ and $\hat{b}=\sqrt{\frac{N}{2}}\sqrt{1+\hat{z}}e^{-i\hat{\theta}/2}$ for annihilation operators $\hat{a}$ and $\hat{b}$, respectively. The quantized Hamiltonian obtained from~\eqref{H_BJJ} for the BJJ-model now has the form 
\begin{equation}\label{A_H_z}
\hat{H}_{BJJ} = \kappa N\left(-\frac{\lambda}{2}\hat{z}^2 - \frac{1}{N}\left(\hat{a}^\dag\hat{b} + \hat{b}^\dag\hat{a}\right)\right).
\end{equation}

For the SJJ-model we start from the classical Hamiltonian~\eqref{H_eff} rewriting it as \begin{equation}\label{H_1} H_{SJJ}=\kappa N\left(-\frac{\Lambda}{2}z^2 - (1-0.21z^2)\,(\sqrt{1-z^2}\cos[\theta])\,\,\sqrt{1-z^2}\right),
\end{equation}
that is suitable for quantization procedure described above. The last term in~\eqref{H_1} we treat by using (formal) Taylor expansion representation for the operator $\sqrt{1-\hat{z}^2}=\sum_{k=0}^\infty C_{1/2}^k(-1)^k\hat{z}^{2k}$. Finally, the quantum version of Hamiltonian~\eqref{H_1} reads as 
\begin{equation}\label{H_2}
\hat{H}_{SJJ}=\kappa N\left(-\frac{\Lambda}{2}\hat{z}^2 - \frac{1}{2N}\left(\sum_{k=0}^\infty C_{1/2}^k(-1)^k(1-0.21\hat{z}^2)\left(\hat{a}^\dag\hat{b}+\hat{b}^\dag\hat{a}\right)\hat{z}^{2k} + H.C. \right)\right).
\end{equation}

Thereafter, we refer to Eq.~\eqref{H_2} as quantum SJJ-model that characterizes the coupled quantum bright soliton problem in a two-mode approximation. Unlike familiar BJJ-model, atom number difference dependent tunneling is also at the core of the quantum SJJ-model, cf.~\eqref{A_H_z}.

Formally, from Eq.~\eqref{H_2} it is clear that $\Lambda$-parameter for the SJJ-model plays the same role as parameter $\lambda$ in the conventional two-mode BJJ-model, see~\eqref{A_H_z}. However, it is possible to see that $\Lambda = \wp\lambda^2\sim N^2$, where we introduced parameter $\wp\equiv\pi\kappa/2\nu$ that characterizes differences in the BJJ- and SJJ-models experimental realization.

\subsection{Macroscopic superposition states in Hartree approximation}

Now we consider the features of macroscopic states for coupled solitons possessing a quite large number of atoms in two-mode approximation. These features can be revealed in the framework of the Hartree approximation applied to solitons~\cite{Haus1989,Alodjants1995} and quantum dimers~\cite{Wright1993}. In particular, the Hartree approximation presumes factorization of $N$-particle state, which allows to represent it as $\left|\psi_N\right\rangle=\left|\psi\right\rangle\otimes\left|\psi\right\rangle\otimes...\left|\psi\right\rangle$, where $\left|\psi\right\rangle $ is a single particle (qubit) state~\cite{Cirac1998,Haus1989}.

To be more specific we are interested in the quantum field variational approach to the model described by~\eqref{H_2} exploiting an atom-coherent state ansatz in the form
\begin{equation}\label{2anz}
 \left|\psi_N\right\rangle = \frac{1}{\sqrt{N!}}\left[\alpha\hat{a}^\dag + \beta\hat{b}^\dag\right]^N\left|0\right\rangle,
\end{equation}
where $\left|0\right\rangle\equiv\left|0\right\rangle_a\left|0\right\rangle_b$ is a two-mode vacuum state and $\alpha$, $\beta$ are unknown (variational) wave functions obeying normalization condition $|\alpha|^2+|\beta|^2=1$. The mean energy in respect of state~\eqref{2anz} is obtained from~\eqref{H_2} and looks like
\begin{equation}\label{2E_1}
E(\alpha,\alpha^*,\beta,\beta^*) = \left\langle\hat{H}_{SJJ}\right\rangle = \kappa N\left(-\frac{\Lambda}{2}S^2 - 2|\alpha||\beta|\left(1 - 0.21S^2\right)\left(\alpha^*\beta + \beta^*\alpha\right)\right),
\end{equation}
where $S\equiv\left\langle\psi_N\Big|\hat{z}\Big|\psi_N\right\rangle=|\beta|^2-|\alpha|^2$.

The energy $E$ in Eq.~\eqref{2E_1} implies Lagrangian, $L$, given by 
\begin{equation}\label{Lag}
L = \frac{i}{2}\left[\alpha^*\dot{\alpha} - \dot{\alpha}^*\alpha + \beta^*\dot{\beta} - \dot{\beta}^*\beta\right]- \kappa N\left(-\frac{\Lambda}{2}S^2 - 2|\alpha||\beta|\left(1 - 0.21S^2\right)\left(\alpha^*\beta + \beta^*\alpha\right)\right).
\end{equation}
By using Eq.~\eqref{Lag} we derive the equations of motion for the field variables $\alpha,\beta$ in the form,
\begin{subequations}\label{shortGP}
\begin{eqnarray}
i\dot{\alpha}&=&\kappa N\Big(\alpha S\Lambda - \alpha\frac{|\beta|}{|\alpha|}\left(1 - 0.21S^2\right)\left(\alpha^*\beta + \beta^*\alpha\right)\\
&-& 0.84\alpha S|\alpha||\beta|\left(\alpha^*\beta + \beta^*\alpha\right) - 2\beta|\alpha||\beta|\left(1 - 0.21S^2\right)\Big);\nonumber\\
i\dot{\beta}&=&\kappa N\Big(-\beta S\Lambda - \beta\frac{|\alpha|}{|\beta|}\left(1 - 0.21S^2\right)\left(\alpha^*\beta + \beta^*\alpha\right)\\
&+& 0.84\beta S|\alpha||\beta|\left(\alpha^*\beta + \beta^*\alpha\right) - 2\alpha|\alpha||\beta|\left(1 - 0.21S^2\right)\Big),\nonumber
\end{eqnarray}
\end{subequations} 
where it is assumed that $|\alpha|\neq0$, or $|\beta|\neq0$ that implies $S\neq\pm1$.

Setting in~\eqref{shortGP} stationary solutions $\alpha(t)\equiv\alpha e^{-i\kappa N\mu t}$; $\beta(t)\equiv\beta e^{-i\kappa N\mu t}$ we get
\begin{subequations}\label{sationary}
\begin{eqnarray}
\mu&=&S\Lambda - \frac{|\beta|}{|\alpha|}\left(1 - 0.21S^2\right)\left(\alpha^*\beta + \beta^*\alpha\right)\\
&-& 0.84S|\alpha||\beta|\left(\alpha^*\beta + \beta^*\alpha\right) - 2\frac{\beta}{\alpha}|\alpha||\beta|\left(1 - 0.21S^2\right)\Big);\nonumber\\
\mu&=& - S\Lambda - \frac{|\alpha|}{|\beta|}\left(1 - 0.21S^2\right)\left(\alpha^*\beta + \beta^*\alpha\right)\\
&+& 0.84S|\alpha||\beta|\left(\alpha^*\beta + \beta^*\alpha\right) - 2\frac{\alpha}{\beta}|\alpha||\beta|\left(1 - 0.21S^2\right)\Big),\nonumber
\end{eqnarray}
\end{subequations} 
Eqs.~\eqref{sationary} lead to a single equation: 
\begin{equation}\label{single}
S\Lambda - S\left(1.42 - 0.63S^2\right)\frac{\alpha^*\beta + \beta^*\alpha}{2|\alpha||\beta|} - \left(\frac{\beta}{\alpha} - \frac{\alpha}{\beta}\right)|\alpha||\beta|\left(1 - 0.21S^2\right) = 0.
\end{equation}

In general, the functions $\alpha$ and $\beta$ in Eq.~\eqref{single} can be represented as $\alpha\equiv|\alpha|e^{i\theta_a}$, $\beta\equiv|\beta|e^{i\theta_b}$, where $\theta=\theta_b-\theta_a$ is phase difference. In particular, Eq.~\eqref{single} reduces to 
\begin{equation}\label{2E_31}
 S\left(\frac{\Lambda}{2} + 1.21 - 0.42S^2\right) = 0
\end{equation}
for $\theta=\pi$, and to 
\begin{equation}\label{2E_32}
 S\left(\frac{\Lambda}{2} - 1.21 + 0.42S^2\right) = 0,
\end{equation}
for $\theta=0$, respectively.
Notice, if we change the sign of tunneling rate, $\kappa$, the shift of phase $\theta$ by $\pi$ is expected for~\eqref{2E_31},~\eqref{2E_32}.

Eqs.~\eqref{2E_31},~\eqref{2E_32} possess a trivial solution with $S\equiv S_0=|\beta_0|^2-|\alpha_0|^2=0$ that is valid for any $\Lambda$. Eq.~\eqref{2E_32} permits two additional solutions, namely solutions $S_\pm=\pm\sqrt{\frac{1}{0.84}\left(2.42 - \Lambda\right)}$ exist for $1.58\leq\Lambda<2.42$, cf.~\eqref{SC_a}. Variables $\alpha$ and $\beta$ which correspond to these solutions are
\begin{subequations}\label{2coefs}
\begin{eqnarray}
&&\alpha_0=\beta_0=\frac{1}{\sqrt{2}};\label{2coefs_a}\\
&&\beta_\pm=\alpha_\mp=\sqrt{\frac{1}{2}\left(1\pm\sqrt{1-X^2}\right)},\label{2coefs_b}
\end{eqnarray}
\end{subequations}
where $X\equiv\sqrt{1-S_\pm^2}=\sqrt{\frac{1}{0.84}\left(\Lambda-1.58\right)}$. 

The energies corresponding to Eqs.~\eqref{2coefs} are given in the form 
\begin{subequations}\label{2E_3}
\begin{eqnarray}
 E_0&=&-\kappa N;\label{2E_3_a}\\
 E_\pm&=&\kappa N\left(0.30\Lambda^2 - 1.44\Lambda + 0.74\right).\label{2E_3_b}
\end{eqnarray}
\end{subequations}
It is important that $E_\pm\geq E_0$ for any $\Lambda$, when $S_\pm$ solution exists. 

Substituting~\eqref{2coefs_b} into~\eqref{2anz} one can obtain ``two halves" of the SC-state
\begin{equation}\label{SC-states}
\left|\psi_{N,\pm}^{(SC)}\right\rangle = \frac{1}{\sqrt{N!}}\left[\alpha_\pm\hat{a}^\dag + \beta_\pm\hat{b}^\dag\right]^N\left|0\right\rangle.
\end{equation}

The states~\eqref{SC-states} can form macroscopic superpositions 
\begin{eqnarray}\label{2cat}
\left|\psi_\pm^{(SC)}\right\rangle&=&\frac{1}{\sqrt{2(1\pm \epsilon)}}\left(\left|\psi_{N,+}^{(SC)}\right\rangle\pm\left|\psi_{N,-}^{(SC)}\right\rangle\right),
\end{eqnarray}
possessing the same energy~\eqref{2E_3_b}.

The states $\left|\psi_\pm^{(SC)}\right\rangle$ in~\eqref{2cat} are mutually orthogonal SC-states defined within the domain $1.58<\Lambda<2.42$, cf.~\eqref{SC_a}. The $\epsilon$-parameter in~\eqref{2cat} characterizes states $\left|\psi_{N,\pm}^{(SC)}\right\rangle$ overlapping (distinguishability) and is defined as
\begin{equation}\label{2cat_size}
\epsilon= \left\langle\psi_{N,+}^{(SC)}|\psi_{N,-}^{(SC)}\right\rangle = X^N.
\end{equation}

At $\epsilon=1$ ($X=1$, $\Lambda\simeq2.42$) the states $\left|\psi_{N,\pm}^{(SC)}\right\rangle$ cannot be distinguished at all. In this case, we have $\beta_\pm=\alpha_\mp=1/\sqrt{2}$ in~\eqref{2coefs_b} and SC-states definition~\eqref{2cat} becomes useless.

In another limit, setting $X\rightarrow 0$ in~\eqref{2cat_size} we obtain $\epsilon\rightarrow 0$ that corresponds to macroscopically distinguishable states 
$\left|\psi_{N,\pm}^{(SC)}\right\rangle$.

As it follows from~\eqref{2cat_size}, at $\epsilon=0$ ($X=0$ ) the states $\left|\psi_{N,\pm}^{(SC)}\right\rangle$ are orthogonal, i.e. maximally distinguishable. In this limit the SC-states approaches the $N00N$-state that occurs for $X=0$ ($\Lambda=1.58$), see~\eqref{2coefs_b},~\eqref{SC-states}, and~\eqref{2cat}. 

In other words, the $N00N$-state appears for $S=\pm1$ that implies $|\alpha|=0$ or $|\beta|=0$ in~\eqref{2anz}-\eqref{Lag} (see also Eqs.~\eqref{N00N}). These solutions are
\begin{subequations}
\begin{eqnarray}\label{2n00n_0}
\left|\psi_{N,-}^{(N00N)}\right\rangle&=&e^{iN\theta_a}\left|N\right\rangle_a\left|0\right\rangle_b = \frac{\left(\hat{a}^\dag\right)^Ne^{iN\theta_a}}{\sqrt{N!}}\left|0\right\rangle;\\
\left|\psi_{N,+}^{(N00N)}\right\rangle&=&e^{iN\theta_b}\left|0\right\rangle_a\left|N\right\rangle_b = \frac{\left(\hat{b}^\dag\right)^Ne^{iN\theta_b}}{\sqrt{N!}}\left|0\right\rangle,
\end{eqnarray}
\end{subequations}
where ``$\pm$" correspond to the sign of $S$ that defines two ``halves" of the $N00N$-state:
\begin{eqnarray}\label{N00N-states}
\left|N00N\right\rangle&=&\frac{1}{\sqrt{2}}\left(\left|\psi_{N,-}^{(N00N)}\right\rangle + e^{iN\theta}\left|\psi_{N,+}^{(N00N)}\right\rangle\right). 
\end{eqnarray}
In~\eqref{N00N-states} we omit the common phase factor $e^{iN\theta_a}$. State~\eqref{N00N-states} possesses phase independent mean energy 
\begin{equation}\label{N00N_energy}
E^{(N00N)} = -\frac{1}{2}\kappa N\Lambda.
\end{equation}

The phase $\theta$ in~\eqref{N00N-states} obey Eq.~\eqref{N00N_b} and defines the domain $0<\Lambda\leq1.58$ for $\Lambda$-parameter, where the $N00N$-state can appear in the framework of the Hartree approach discussed above.

Noteworthy, the state (35) is fragile enough and rapidly collapses to states $\left|N-1\right\rangle_a\left|0\right\rangle_b$ or $\left|0\right\rangle_a\left|N-1\right\rangle_b$ even in the presence of one particle loss, cf.~\cite{Vogel2015}. As we see below, the SJJ-model exhibits some new specifics for the $N00N$-state formation in a purely quantum limit. 

\section{Mesoscopic quantum superposition states}

\subsection{Quantum energy spectrum}

Now let us consider purely quantum features of the superposition states for the SJJ-model beyond the Hartree approach. This approximation presumes a quite large number of particles, that can be thousands or more in experiments with atomic condensates, see e.g.~\cite{Colzi2018}. However, for the condensates possessing the negative scattering length the number of atoms is limited due to the collapsing effect. For the BECs confined in optical lattices and representing familiar Bose-Hubbard system the number of atoms varies from several tens to several hundreds per site, see e.g.~\cite{Lewenstein2007}, representing a mesoscopic system with a moderate number of particles. In respect of the number of particles, mesoscopic (in Greek ``meso" means ``middle") regime is located between microscopic (up to tens of particles) and macroscopic (thousand particles and more) regimes and represents a great interest for current studies in quantum physics~\cite{Arndt2009}. The formation of efficient quantum mesoscopic superposition states represents one of challenging problems both in theory and experiment~\cite{Sinatra2017}. It is worth mentioning that the crossover from a coherent to Fock (or number squeezed) regime possessing some important quantum features plays an important role in this case~\cite{Greiner2002}. In particular, variation of the tunnelling rate between the lattice sites allows to achieve the crossover from a coherent to Fock regime (Mott-insulator state). However, as it is shown below the (self)tuning of this crossover may be much more efficient for the SJJ-model that represents the simplest (two-site) Bose-Hubbard system.

Here we are interested in mesoscopic limit for the SJJ-model illustrated in Fig.~\ref{FIG:W-pot}. We assume that the total number of atoms, $N$, is not too large, and the characteristic temperatures, $T$, $T_c$, relevant to the condensates are sufficiently low in comparison with the characteristic energy space, $\delta E$, between the neighboring low-energy quantum levels of the system~\cite{Andreev1998}; $T_c$ is the critical temperature of the phase transition to the BEC state. Physically it implies the importance how a small number of atoms behave in the presence of a large total number, $N$. In such a limit the analysis of the two-mode system described by Hamiltonians~\eqref{A_H_z},~\eqref{H_2} has to be fully quantum. 

We represent the quantum two-mode state, $|\Psi(\tau)\rangle$, in the Fock basis as (cf.~\cite{Olsen2010}):
\begin{equation}\label{ground_state}
|\Psi(\tau)\rangle = \sum_{n=0}^N A_n(\tau)|N-n,n\rangle,
\end{equation}
where $|N-n,n\rangle\equiv|N-n\rangle_a|n\rangle_b$ denotes the two-mode atom-number state; the coefficients $A_n(\tau)$ obey normalization condition $\sum_{n=0}^N |A_n(\tau)|^2=1$. The $A_n(\tau)$ fulfills the Schr{\"o}dinger equation 
\begin{equation}\label{SE}
i\frac{d A_n(\tau)}{d\tau} = \langle N-n,n|\hat{H}|\Psi(\tau)\rangle.
\end{equation}

In connection with quantum Hamiltonians~\eqref{A_H_z},~\eqref{H_2} the Eq.~\eqref{SE} can be represented as:
\begin{eqnarray}\label{eq_short}
&&i\dot{A}_n = a_nA_{n+1} +\beta_{n-1}A_{n-1},
\end{eqnarray}
where the following notations are introduced:
\begin{subequations}
\begin{eqnarray}\label{A_coeff}
\alpha_n&=&-\frac{\lambda}{2}\left(\frac{2n}{N}-1\right)^2,\\
\beta_n&=&-\frac{1}{N}\sqrt{(n+1)(N-n)}
\end{eqnarray}
\end{subequations}
for the BJJ-model, and
\begin{subequations}
\begin{eqnarray}\label{coeff}
\alpha_n&=&-\frac{\Lambda}{2}\left(\frac{2n}{N}-1\right)^2,\nonumber\\
\beta_n&=&-\frac{1}{N^2}\Bigg(\left[1-0.21\left(\frac{2n}{N}-1\right)^2\right](n+1)\sqrt{(N-n)(N-n-1)}\\
&+&\left[1-0.21\left(\frac{2(n+1)}{N}-1\right)^2\right](N-n)\sqrt{n(n+1)}\Bigg)\nonumber
\end{eqnarray}
\end{subequations}
for the SJJ one.

Then let us focus on stationary solution $A_n(\tau)=A_ne^{-iE_n\tau}$ of~\eqref{eq_short} that enables to find out energy spectrum ${E_n}$ of the quantum Hamiltonian~\eqref{H_2} for the SJJ-model.

\begin{figure}[t]
\center{\includegraphics[width=\linewidth]{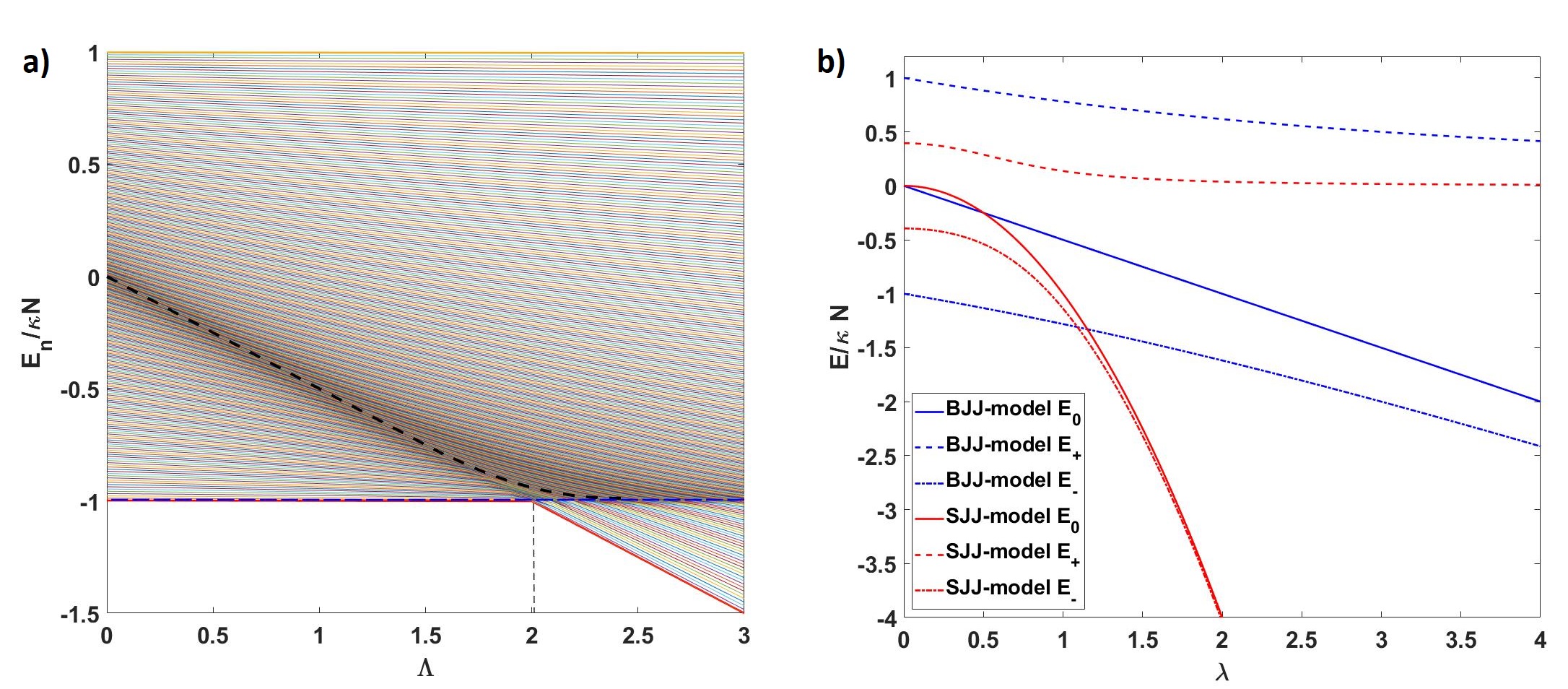}}
\caption{(a) Dependencies of the quantum Hamiltonian $\hat{H}$ eigenenergies, $E/\kappa N$, vs. $\Lambda$-parameter for $N=300$ particles. Black and blue dashed lines characterize mean energies in Hartree approximation. Quantum energy spectrum is bounded by values $E_0$ (solid red line) and $E_{300}$ (upper solid yellow line). Vertical dashed line indicates the critical value $\Lambda_c\approx2.0009925$ of vital parameter $\Lambda$ for the SJJ-model. (b) Energy spectrum for $N=2$ particles vs. $\lambda$-parameter. The key parameter $\Lambda$ is equal to $\wp\lambda^2$, $\wp=2$ for the SJJ-model. Red and blue lines correspond to Eqs.~\eqref{3_l_1} and~\eqref{3_l_2}, respectively.}
\label{FIG:spectrum}
\end{figure}

Fig.~\ref{FIG:spectrum}(a) demonstrates numerical calculations of the spectrum for $N=300$ particles as a function of $\Lambda$-parameter. Practically, the dependence on $\Lambda$ can be examined with the help of a Feshbach resonance through controlling atom-atom scattering length ($u$-parameter)~\cite{Malomed2006,Cheng2010}. 

The black dashed curve in Fig.~\ref{FIG:spectrum}(a) indicates the steady-state solutions for the SJJ-model in the energy dense region, where the Hartree approach is satisfied, cf.~\cite{Wright1993}; this curve characterizes the energies~\eqref{N00N_energy} and~\eqref{2E_3_b}, and is valid for $0<\Lambda\leq1.58$ and $1.58<\Lambda\leq2.42$, respectively. The blue dashed line corresponds to the energy $E_0$ in accordance with Eq.~\eqref{2E_3_a}. 

Main peculiarities with the coupled soliton energies occur in the vicinity of point $\Lambda\equiv\Lambda_c\approx 2.0009925$. In particular, at $\Lambda=\Lambda_c$ the SJJ-system loses its coherence features entering Fock (Mott-insulator) regime modifying particle number fluctuations, cf.~\cite{Leggett2001,Salasnich2011,Gati2007}. Notice, that in this case, the SJJ-system treated classically changes its regime from Rabi-like oscillations to self-trapping, cf.~\cite{Raghavan1999,Tsarev2018,Raghavan2000}. In particular, at $\Lambda<\Lambda_c$ energetically favorable state for the quantum SJJ-system in Fig.~\ref{FIG:spectrum}(a) is described by the blue dashed (or the red solid) line exhibiting the minimal energy $E_0$. As seen from Fig.~\ref{FIG:spectrum}(a), at $\Lambda\geq\Lambda_c$ the energies $E_\pm$, $E^{(N00N)}$, which correspond to the superposition states in the Hartree approximation, are comparable with $E_0$ but placed a little bit above it.
 
Situation changes completely in a purely quantum limit when parameter $\Lambda$ crosses critical its value $\Lambda_c$. The red solid line in Fig.~\ref{FIG:spectrum}(a) clearly shows that the quantum $N00N$-state is energetically favorable at $\Lambda\geq\Lambda_c$ for the SJJ ground state (it is demonstrated below that this state is not exactly a $N00N$-state as it happens in the semiclassical limit).

To understand the main properties of the SJJ-model at low energies ($n=0$) we examine the simplest case with $N=2$ particles. The diagonalization of~\eqref{H_2} gives three energy eigenvalues obtained analytically:
\begin{subequations}\label{3_l_1}
\begin{eqnarray}
E_0&=&-\kappa N\frac{\Lambda}{2};\\
E_\pm&=&\frac{1}{4}\kappa N\left[-\Lambda\pm\sqrt{\Lambda^2+2.5}\right].
\end{eqnarray}
\end{subequations}

Similar results may be obtained for the BJJ-model. Diagonalization of the relevant Hamiltonian~\eqref{A_H_z} leads to 
\begin{subequations}\label{3_l_2}
\begin{eqnarray}
E_0&=&-\kappa N\frac{\lambda}{2};\\
E_\pm&=&\frac{1}{4}\kappa N\left[-\lambda\pm\sqrt{\lambda^2+16}\right].
\end{eqnarray}
\end{subequations}
The energy eigenvalues~\eqref{3_l_1} and~\eqref{3_l_2} are plotted in Fig~\ref{FIG:spectrum}(b) as functions of $\lambda$ that is relevant to the key parameter combination $uN/\kappa$ and inherent to both JJ-models. The gap $\Delta E=E_0-E_-$ between two lowest energy eigenstates obtained from Eqs.~\eqref{3_l_1},~\eqref{3_l_2} looks like 
\begin{subequations}\label{Delta}
\begin{eqnarray}
\Delta E_{SJJ}&=&\frac{1}{4}\kappa N\left[-\Lambda +\sqrt{\Lambda^2+2.5}\right];\\
\Delta E_{BJJ}&=&\frac{1}{4}\kappa N\left[-\lambda +\sqrt{\lambda^2+16}\right].
\end{eqnarray}
\end{subequations}

From Eqs.~\eqref{Delta} it follows $\Delta E_{SJJ}\simeq 0.4\kappa N$ and $\Delta E_{BJJ}=4\kappa N$ for vanishing nonlinearity $u\approx0$. Notably, the limiting case with $u=0$ is not valid for the SJJ-model since bright solitons do not exist in the absence of nonlinearity. On the contrary, for the relatively large nonlinear parameter, such as $\lambda,\Lambda\gg 1$, from~\eqref{Delta} we obtain for the energy gaps $\Delta E_{BJJ}\approx4\kappa^2/u$, and $\Delta E_{SJJ}\approx5\kappa^2/Nu^2\approx5\Delta E_{BJJ}/4Nu$, respectively. The suppression of this gap with increasing atom-atom scattering length ($u$-parameter) is clearly seen in Fig~\ref{FIG:spectrum}(b). 

For the moderate values of $\lambda$ and $\Lambda$ parameters $\Delta E_{SJJ}$ is $N$ times larger than $\Delta E_{BJJ}$. The lifetime $T_{E_{-}}$ for transition $\Delta E_{-}$ is proportional to $\hbar/\Delta E_{-}$, see e.g.~\cite{Messiah1968}. It is evident that in general even for a small number of particles the lifetime of soliton junctions may be sufficiently larger than for the conventional BJJ-model, cf.~\cite{Wright1993}.

\begin{figure}[t]
\center{\includegraphics[width=\linewidth]{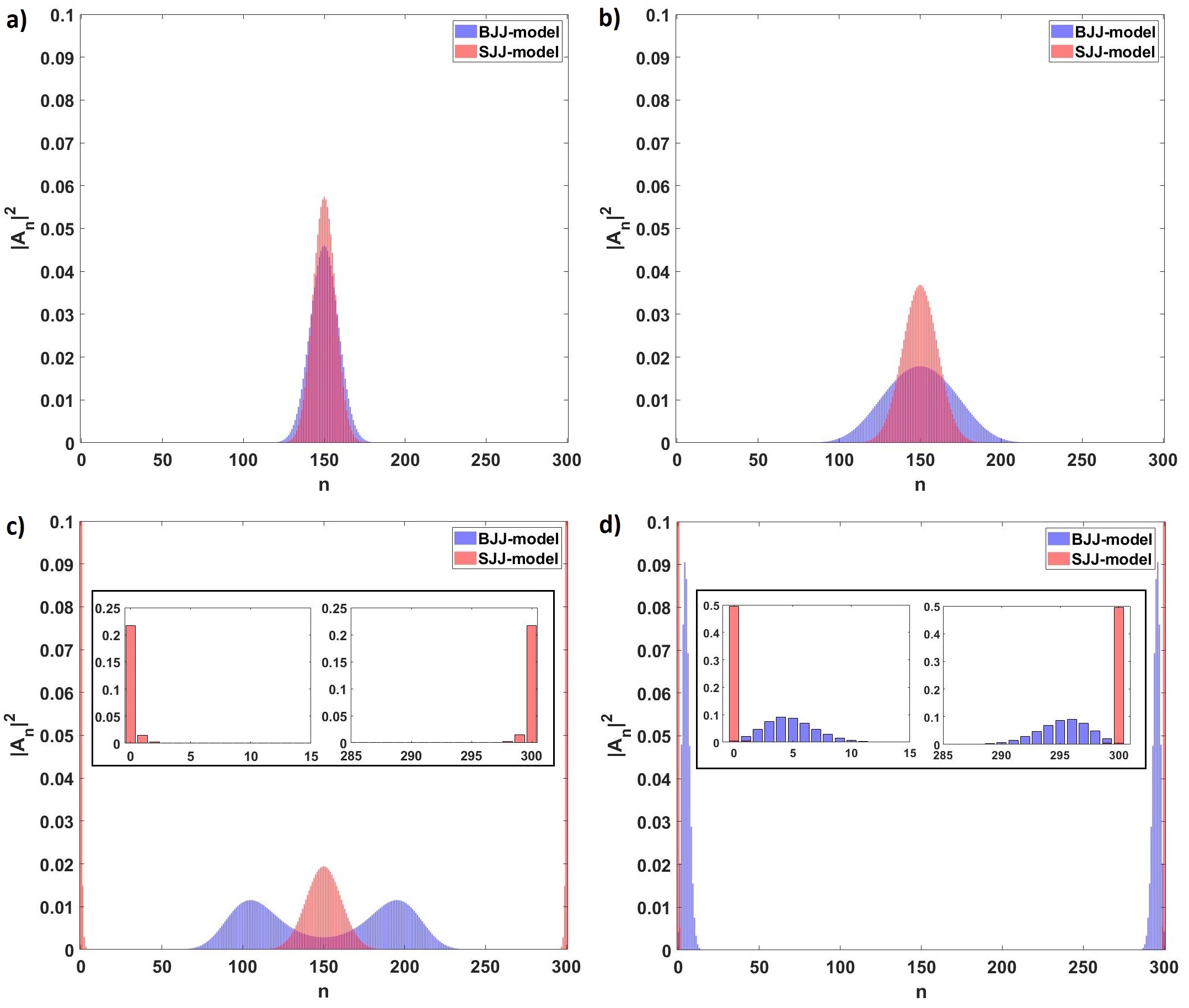}}
\caption{Ground state probabilities $|A_n|^2$ vs. $n$ for (a) $\Lambda=0$, $\lambda=0$; (b) $\Lambda=2$, $\lambda=1$; (c) $\Lambda\approx 2.0009925$, $\lambda=1.06$; (d) $\Lambda=\lambda = 4$. The total particle number is $N=300$. Insets in (c,d) show the probabilities in the vicinity of ``edge" states with $n=0$ (left-hand panels) and $N=300$ (right-hand panels), respectively.}
\label{FIG:ground_states}
\end{figure}

In Fig.~\ref{FIG:ground_states} we represent the ground state probability distributions, $|A_n|^2$, (eigenfunctions of~\eqref{H_2} with the minimal energies) for different values of the key parameters $\Lambda$ and $\lambda$ for the SJJ- and BJJ-models, respectively. The plots for the SJJ-model are established by red and can be easily compared with the ones for the BJJ-model indicated by blue. Fig.~\ref{FIG:ground_states}(a) demonstrates a small difference in distributions for the SJJ- and BJJ-models in an ideal (non-interacting) atomic gas limit. Since the total number of particles is large enough, the initial distribution approaches Gaussian for the BJJ-model. 

The differences between distributions considered for the SJJ- and BJJ-models become evident and important in Fig.~\ref{FIG:ground_states}(b)-(d) due to the particle number dependence of the effective tunneling rate, $\kappa_{eff}$, see~\eqref{Lambda}. Figures~\ref{FIG:ground_states}(b) and~\ref{FIG:ground_states}(c) exhibit features of examined JJ-systems immediately before and after the crossover, respectively. In the vicinity of the critical value $\Lambda_{c}\simeq2$ the distribution for the SJJ-model becomes significantly non-Gaussian, see Fig.~\ref{FIG:ground_states}(b). 

The SJJ- and BJJ-models behave differently at the crossover region for parameters $\Lambda\geq\Lambda_{c}$ and $\lambda\geq\lambda_{c}$, respectively. In particular, deepening in ground state probability $|A_n|^2$ takes place for the BJJ-system with increasing $\lambda$. 

Behavioral scenarios after crossover points $\Lambda_c\simeq2$ and $\lambda_c\simeq1$ for the SJJ- and BJJ-models, respectively, are represented in Fig.~\ref{FIG:ground_states}(c,d). In particular, two probability peaks corresponding to the macroscopic SC-state arise for BJJs. Further increasing of $\lambda$ results in the growth of the ``distance" between the cats ``moving" to the ``edges", $n=0$ and $n=N$, see Fig.~\ref{FIG:ground_states}(d). The $N00N$-state,
\begin{equation}\label{N00N2}
\left|N00N\right\rangle = \frac{1}{\sqrt{2}}\left(\left|N\right\rangle_a\left|0\right\rangle_b + e^{-i\theta_N}\left|0\right\rangle_a\left|N\right\rangle_b\right),
\end{equation}
occurs for the BJJ-model in the limit of a very large $\lambda$-parameter, see below Fig.~\ref{FIG:HZ}(b) and cf.~\cite{Olsen2010,Salasnich2011}. 

In general, the SJJ-model demonstrates evident advantages to achieve a $N00N$-state. Roughly speaking, state (37) for the SJJ-model after the crossover represents a superposition of entangled Fock states possessing a large $N00N$-state component - see. Fig.~\ref{FIG:ground_states}(c). A physical explanation for this phenomenon is illustrated as follows. 

First, for a given tunneling rate, $\kappa$, the $\Lambda$-parameter depends on $N^2$, which allows to achieve a required crossover point for the moderate number of particles, $N$.

Second, as it follows from Fig.~\ref{FIG:ground_states}(b,c) the crossover for the SJJ-model happens abruptly with a slightly increasing $\Lambda$ parameter. The dependence of the effective tunneling rate, $\kappa_{eff}$, on the atom number leads to a significant improvement of the ``edge" states occupation. In particular, as clearly seen in Fig.~\ref{FIG:ground_states}(c), two large peaks of probability distribution occur at the ``edges" with $n=0$ and $n=N$, respectively. They correspond to the $N00N$-state. Simultaneously, there exist small (non-vanishing symmetric) peaks for $n=1, N-1$, $n=2, N-2$, etc., in the vicinity of ``edge" states - see the inset in Fig.~\ref{FIG:ground_states}(c).

Furthermore, the increment in $\Lambda$ leads to the suppression of small peaks probabilities, but to the enhancement of the ``edge" states $n=0, N$ population. The distribution approaches approximately ``ideal" $N00N$-state~\eqref{N00N2} at moderate values of $\Lambda$, cf. Fig.~\ref{FIG:ground_states}(d). However, as we show below, the presence of small components in the satellite entangled Fock state plays an important role for the resistance of state~\eqref{ground_state} to weak particle losses, cf.~\cite{Huver2008}.

\subsection{Entanglement, steering and planar spin-squeezing}

\begin{figure}[t]
\center{\includegraphics[width=\linewidth]{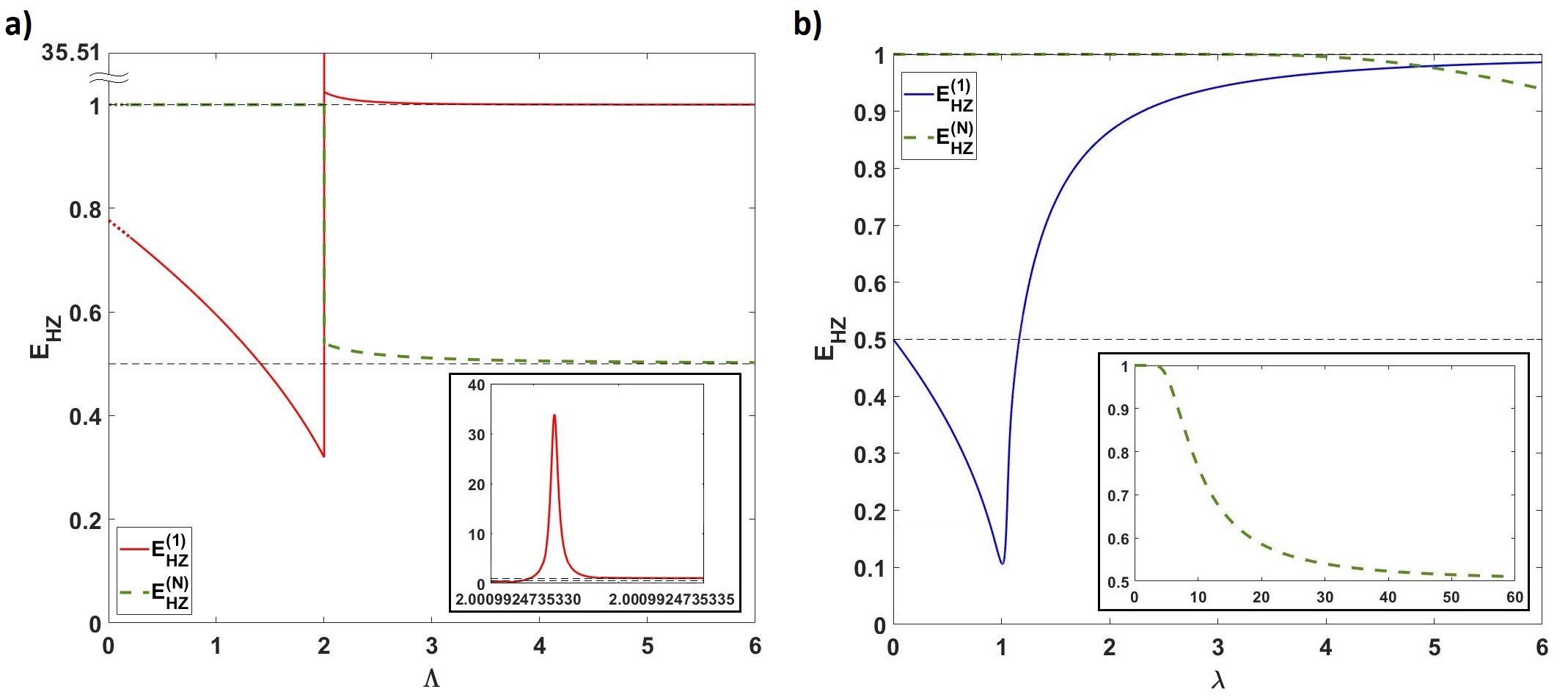}}
\caption{$E_{HZ}^{(1,N)}$ for (a) the SJJ- and (b) BJJ-models as functions of parameters $\Lambda$ and $\lambda$, respectively. 
The inset in (a) demonstrates the behavior of $E^{(1)}_{HZ}$ within a tiny crossover region. The inset in (b) demonstrates $E_{HZ}^{(N)}$ at large $\lambda$.} 
\label{FIG:HZ}
\end{figure}

The complete analysis of the highly nonclassical states formation requires investigating high (two- and much higher) order correlation functions. Entanglement and spin squeezing are important prerequisits of non-classical behavior of multiparticle atomic system. To be more specific we examine here so-called $m$-order Hillery-Zubairy (HZ) criterion~\cite{Reid2011,Reid2018} defining as 
\begin{equation}\label{HZ_m}
E_{HZ}^{(m)} = 1 + \frac{\left\langle\hat{a}^{\dag m}\hat{a}^m\hat{b}^{\dag m}\hat{b}^m\right\rangle - \left|\left\langle\hat{a}^m\hat{b}^{\dag m}\right\rangle\right|^2}{\left\langle\hat{a}^{\dag m}\hat{a}^m\left(\hat{b}^m\hat{b}^{\dag m} - \hat{b}^{\dag m}\hat{b}^m\right)\right\rangle}.
\end{equation}
The entanglement occurs when inequalities 
\begin{equation}\label{Ineq}
0\leq E_{HZ}^{(m)}<1
\end{equation}
are satisfied. The value $E_{HZ}^{(m)}=1$ should be discussed separately for some specific states. We examine two limiting cases for definition~\eqref{HZ_m}. In particular, for $m=1$ we can recast Eq.~\eqref{HZ_m} in terms of $N$-particle spin-operators as following:
\begin{equation}\label{HZ_spin}
E^{(1)}_{HZ}=\frac{\delta J_{X}+\delta J_{Y}}{\langle\hat{J}\rangle},
\end{equation}
where $\delta J_{j}\equiv \left\langle\left(\Delta\hat{J}_j\right)^2\right\rangle = \left\langle\hat{J}_j^2\right\rangle - \left\langle\hat{J}_j\right\rangle^2$ is the variance of fluctuations for $j$-th component of spin, established by operators 
\begin{center}
\begin{eqnarray}
&&\hat{J}_X = \frac{1}{2}\left(\hat{a}^\dag\hat{b} + \hat{b}^\dag\hat{a}\right);\textrm{ }\hat{J}_Y = \frac{1}{2i}\left(\hat{b}^\dag\hat{a} - \hat{a}^\dag\hat{b}\right);\textrm{ }\hat{J}_Z = \frac{1}{2}\left(\hat{a}^\dag\hat{a} - \hat{b}^\dag\hat{b}\right);\nonumber\\
&&\hat{J} = \frac{1}{2}\left(\hat{a}^\dag\hat{a} + \hat{b}^\dag\hat{b}\right) = \frac{1}{2}\hat{N}.
\end{eqnarray}
\end{center}
Combining Eq.~\eqref{HZ_m} with~\eqref{Ineq} one can see that strong entanglement can be obtained if variances $\delta J_{X}$ and $\delta J_{Y}$ are minimal for a given atom number, $N$. Strictly speaking, $\delta J_{\parallel}=\delta J_{X}+\delta J_{Y}$ represents the variance of fluctuations of atomic spin in $XY$-plane. Red curve in Fig.~\ref{FIG:HZ}(a) demonstrates behavior of $E_{HZ}^{(1)}$ versus vital parameter $\Lambda$. Since the SJJ-model is invalid for ideal atomic gases, we represent a relevant curve dashed in the vicinity of $u=0$. For the BJJ-model (see blue curve in Fig.~\ref{FIG:HZ}(b)) in this limit we can exploit the two-mode condensate state like~\eqref{2anz} with $|\alpha|^2=|\beta|^2=0.5$ that gives $E_{HZ}^{(1)}=0.5$. 

The EPR steering entanglement is achieved in the region, where $E_{HZ}^{(1)}<0.5$, cf.~\cite{Reid2012, Jones2007}. From Fig.~\ref{FIG:HZ}(a) it is evident that this area occurs close to the point $\Lambda_c\simeq2$. Noteworthy, the sharp behavior of $E_{HZ}^{(1)}$ within the crossover region appear in the absence of losses, see the inset in Fig.~\ref{FIG:HZ}(a). Evidently, broadening of this region is expected if one- and, especially, three-body losses are taken into account.

The minimal value $E_{HZ, min}^{(1)}$ of parameter $E_{HZ}^{(1)}$, that we denote as $C_J$, is determined through uncertainties in $J_X$ and $J_Y$ components which do not commute. The $C_J$-coefficient is estimated from inequality 
\begin{equation}\label{C-limit}
E_{HZ}^{(1)}\geq C_{J}.
\end{equation}
 
\begin{figure}[t]
\center{\includegraphics[width=0.5\linewidth]{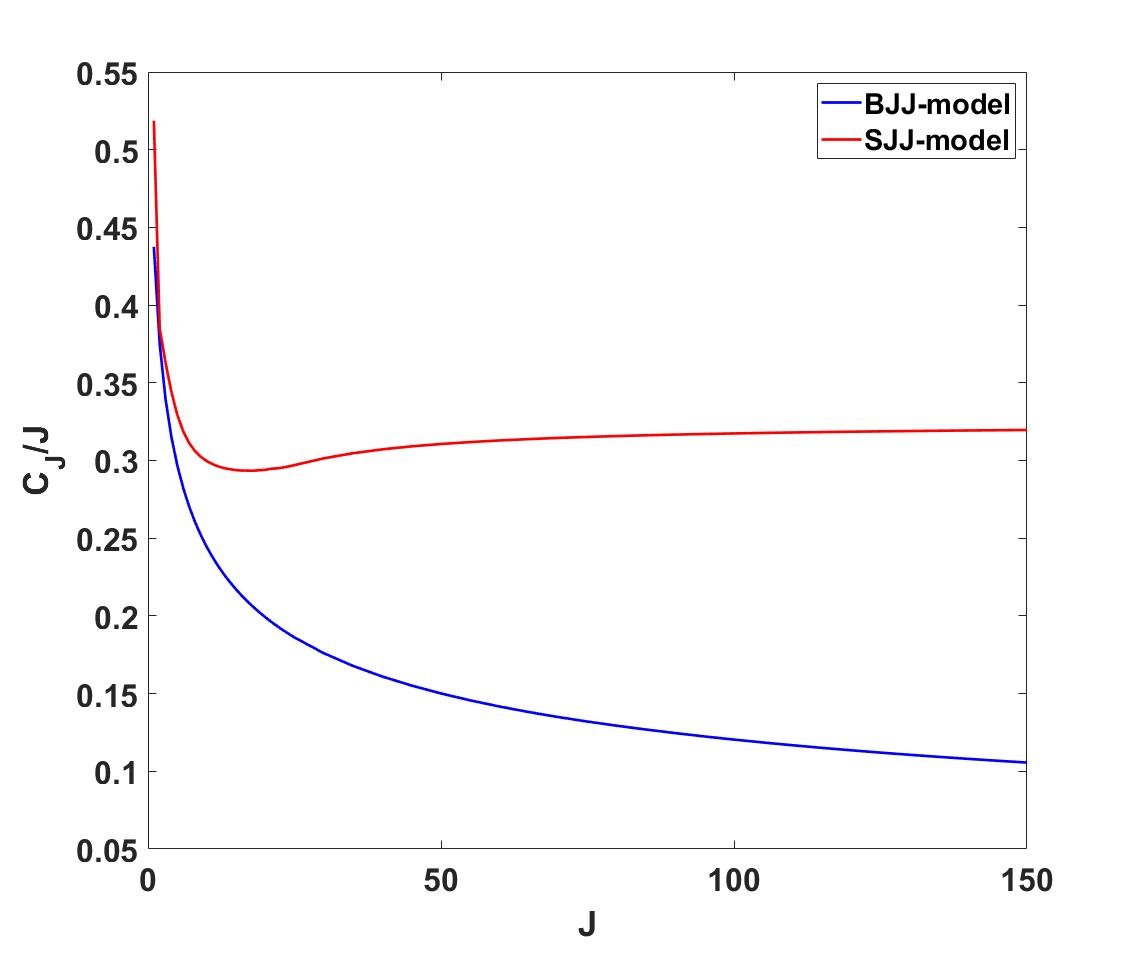}}
\caption{Normalized coefficient $C_J$ vs. $J=N/2$ for the SJJ- and BJJ-models.}
\label{FIG:C-limit}
\end{figure}

In Fig.~\ref{FIG:C-limit} we plot the parametric dependence of the normalized $C_J$-coefficient on a spin vector ``length", $J$, that is simply a half of the particle number, $N/2$. In particular, the BJJ-model exhibits the largest depth of steering determined by $C_J$. 

The coefficient $C_J$ theoretically can be estimated by using the Gaussian distribution function possessing width, $\sigma$, (see Fig.~\ref{FIG:ground_states}(b)), cf.~\cite{Reid2011}. In particular, for the BJJ-model $\sigma_{BJJ}=(2J)^{-2/3}$ and $C_J^{(BJJ)}\propto 1/\sigma_{BJJ}=0.6J^{-1/3}$ giving the minimum of the blue curve in Fig.~\ref{FIG:HZ}(b) about $0.163$. The magnitude $C_J^{(SJJ)}$ (that is about $0.31924$) for the SJJ-model in Fig.~\ref{FIG:HZ}(a) (red curve) is not so simple to estimate analytically because of essentially nonlinear behavior happening close to the value $\Lambda_c\simeq2$ (or, $\lambda_c\simeq1$). However, as clearly seen from Fig.~\ref{FIG:ground_states}(b) the distribution for the SJJ-model is sufficiently narrower than for the BJJ one, i.e. $\sigma_{SJJ}<\sigma_{BJJ}$, and we expect $C_J^{(SJJ)}>C_J^{(BJJ)}$ to take place for curves exhibiting $E_{HZ}^{(1)}$ in Fig.~\ref{FIG:HZ}(a,b). 

Notably, the suppression of variance $\delta J_{\parallel}$ manifests planar squeezing defined as (cf.~\cite{Puentes2013}) 
\begin{equation}\label{Squeezing}
\delta J_{\parallel}< J_{\parallel},
\end{equation}
where $J_{\parallel}\equiv\sqrt{\left\langle\hat{J}_X\right\rangle^2 + \left\langle\hat{J}_Y\right\rangle^2}$ is the magnitude of the spin component in $XY$-plane. In particular, for both SJJ- and BJJ-models we have $\left\langle\hat{J}_X\right\rangle\approx J$, $\left\langle\hat{J}_Y\right\rangle=\left\langle\hat{J}_Z\right\rangle=0$ within the domain $0<\Lambda \leq 2$ for Fig.~\ref{FIG:HZ}(a) and $0\leq \lambda \leq1$ for Fig.~\ref{FIG:HZ}(b), respectively. Thus, in this area $E_{HZ}^{(1)}$ reflects the planar spin squeezing that occurs due to significant enhancement of fluctuations in the $Z$-component of spin~\cite{Reid2011}. 

It is important to stress that in Fig.~\ref{FIG:HZ}(a) (green dashed line) the SJJ-system undergoes the abrupt transition to $N00N$-state~\eqref{N00N2} at the crossover region for $\Lambda_c\simeq2$ that is consistent with Fig.~\ref{FIG:ground_states}(c). The $E_{HZ}^{(m)}$-parameter calculated for $m=1$ and $m=N$ by using $N00N$-state~\eqref{N00N2} gives $E_{HZ}^{(1)}=1$, $E_{HZ}^{(m)}=0.5$, respectively. 

From Fig.~\ref{FIG:HZ}(b) (green dashed line) it is evident that the transition to the $N00N$-state for the BJJ-model occurs upon the large scales of $\lambda>>1$. 

\subsection {Quantum state engineering in the presence of losses}

We now outline the possibility of experimental observation of $N00N$-states with solitons with feasible condensate parameters. To be more specific we discuss lithium condensate solitons~\cite{Khaykovich2002}. In addition, we would like to mention the work~\cite{Nguyen2014} where authors recently studied lithium atomic soliton collisions on relatively long times, i.e. on $32ms$ and more. In particular, we focus on two weakly coupled condensates of $N=300$ atoms ($|a_{sc}|\simeq1.4\,nm$, $uN\sim1.88$), confined in an asymmetric trap as it is shown in Fig.~\ref{FIG:W-pot}, with $\omega_x/2\pi=70\,\, Hz$ and radial trapping frequency $\omega_{\perp}/2\pi=700\,\, Hz$ that imposes $a_{\perp}\simeq 1.4\,\, \mu m$. The solitons in~\cite{Khaykovich2002,Nguyen2014} observed within characteristic time $8\div32ms$. The $\Lambda$ parameter achieves its critical value, $\Lambda_c\simeq2$, at tunneling rate $|{\cal K}|/2\pi\simeq77\,\, Hz$ or $|{\cal K}|\simeq3.7\,\,nK$, which correlates the current experiments with atomic JJs, see e.g.~\cite{Gati2007}. Notice, such a tunneling rate promotes effective coupling of solitons within their observation time. Some moderate tuning of $\cal K$ or scattering length $|a_{sc}|$ allows to obtain the crossover to the $N00N$-state with $\Lambda\geq\Lambda_c$.
 
For the BJJ-model the critical value $\lambda_c=1$ is achieved with $|{\cal K}|/2\pi\simeq 83\,\, Hz$. However, the $N00N$-state requires essential enhancement of $\lambda$ in the experiment ($60$ times and more, see inset in Fig.~\ref{FIG:HZ}(b)) that may require a non-trivial experimental task. On the other hand, SC-states can be observed for the BJJ-model with $\lambda\geq 1$, see Fig.~\ref{FIG:ground_states}(d) and~\cite{Sinatra2017}.

Let us briefly discuss the loss issues for the SJJ-model described in this work. This problem is important for applications of the SJJ-model in current quantum technologies, especially in quantum metrology, where $N00N$-states play an essential role, cf.~\cite{Dowling2008,Vogel2015,Kapale2007}. 

Typically, it is instructive to consider one- and three-body losses occurring in condensates~\cite{Knoop2012}. As shown in~\cite{Li2008}, the role of various losses becomes critical at a relatively large number of particles in the condensate. For example, in~\cite{Li2008} it is shown that spin squeezing degrades for a bimodal condensate with $N\geq 10^4$ atoms. However, the $N00N$-state formation in the presence of weak losses requires some special attention in this work. 

First, let us estimate possible losses for the SJJ-model as a prerequisite for $N00N$-state formation. Since vital parameter, $\Lambda$, is proportional to $N^2$, many-body, especially three-body, inelastic collisions bringing to losses may be important~\cite{Fedichev1996}.

Physically three-body losses occur as a result of three-body recombination process appearing with condensate atoms. Such processes primarily remove condensate particles possessing low kinetic energy and those located at the trap center. The three-body losses play a very important role in the vicinity of Feshbach resonance where scattering length essentially enhances~\cite{Grimm2003}.

As with one-body losses we expect here to have dynamical adiabatic changing of regimes from Fock to Josephson or Rabi, if losses are weak~\cite{Sols2002}. On the other hand, since effective tunneling parameter, $\kappa_{eff}$, depends on the particle number, the SJJ-model in the presence of three-body losses may be also relevant to some kind of dissipative tunneling problem~\cite{Caldeira1983}. 

In zero temperature limit the three-body recombination rate is $L_3\propto\hbar a_{sc}^4/m$ that approaches to $L_3\simeq 2.6\times10^{-28}\, cm^{6}/s$ for lithium atoms, see e.g.~\cite{Verhaar1996}. A simple rate equation for particle number leads to decay of condensate atoms with the law as
\begin{equation}\label{3body}
N(t) \simeq \frac{N(0)}{\sqrt{1+2L_3\rho^2t}},
\end{equation} 
where $N(0)$ is initial atom number in the condensate, $\rho$ is atom number density that we assume here as a constant. Taking it as $\rho\simeq10^{13}cm^{-3}$ for the effective rate of three-body losses $\gamma_3=2L_3\,\rho^2$ we obtain $\gamma_3=5.2\times10^{-2}s^{-1}$. 
 
The one-body losses rate may be estimated as $\gamma_1=0.2\, s^{-1}$, see e.g.~\cite{Sinatra2017}. At times of soliton observation in~\cite{Khaykovich2002, Nguyen2014} we effectively obtain $\gamma_{1,3}t<<1$ that allows to examine a non-perturbed SJJ-system. Notice, that in~\cite{Khaykovich2002} observation time was less than $10\,ms$.

This conclusion is supported by recent experiments performed in a more realistic case at finite temperatures~\cite{Gross2009}. In particular, as it is shown in~\cite{Petrov2013}, three-body losses for lithium condensate atoms with scattering length $|a_{sc}|\simeq10.6\,nm$ and tuned by magnetic field at the temperature $T=5.2\,\, \mu K$ are important at the time scales $t\simeq100\,ms$ and more. Especially at times $t\leq32\,ms$ relevant to matter-wave soliton observations in~\cite{Nguyen2014} the particle number reduction is practically negligible. 

The analysis represented above is valid in the mean-field limit. In this sense, it is instructive to examine few particle losses in a purely quantum way for state~\eqref{ground_state}. We can consider an approach to the losses, which is typically used in the framework of $N00N$-state applicability for quantum metrology tasks, see e.g.~\cite{Vogel2015, Huver2008, Dobrzanski2009, Rubin2007}. Various strategies are proposed to achieve maximally accessible scaling for phase estimation; they manifest that in the presence of losses the ideal (balanced) $N00N$-state may be non-optimal. It is shown that entangled Fock states, unbalanced $N00N$-states, and some specific two-component states may be more useful in the presence of different level of losses~\cite{Dorner2009,Huver2008, Dobrzanski2009}. Here we model the losses of condensate particles by means of the fictitious beam splitter (BS) approach. This approach represents a useful tool to study coupling of quantum macroscopic superposition states with environment - see Fig.6. The ``input" two-mode Fock state~\eqref{ground_state} after two BSs transforms into (cf.~\cite{Dobrzanski2009}) 
\begin{equation}\label{bimodal}
\left|N-n\right\rangle_a\left|n\right\rangle_b\rightarrow\sum_{l_b=0}^N\sum_{l_a=0}^{N-l_b}\sqrt{B_{l_a,l_b}^n}\left|N-n-l_a\right\rangle_a\left|n-l_b\right\rangle_b\left|l_a\right\rangle\left|l_b\right\rangle,
\end{equation}
where $l_{a}$ and $l_{b}$ are the numbers of particles lost from ``a" and ``b" wells, respectively. In~\eqref{bimodal} we introduce a coefficient 
\begin{equation}\label{B}
B_{l_a,l_b}^n = \left(
\begin{array}{c}
N-n\\
l_a
\end{array}
\right)\left(
\begin{array}{c}
n\\
l_b
\end{array}
\right)
\eta_a^{N-n}(\eta_a^{-1}-1)^{l_a}\eta_b^n(\eta_b^{-1}-1)^{l_b}.
\end{equation}
that characterizes the relevant probabilities in the presence of particle losses. Here $\eta_a$ and $\eta_b$ ($\eta_{a,b} \leq1$) are the transmissivities of BSs in the channels ``a" and ``b", respectively. Then state $\left|\Psi_{out}\right\rangle$ after particle losses reads as (see Fig.6) 
\begin{equation}\label{out}
\left|\Psi_{out}\right\rangle =\frac{1}{\sqrt{p}} \sum_{l_b=0}^N\sum_{l_a=0}^{N-l_b}\sum_{n=l_b}^{N-l_a}A_n\sqrt{B_{l_a,l_b}^n}\left|N-n-l_a\right\rangle_a\left|n-l_b\right\rangle_b\left|l_a\right\rangle\left|l_b\right\rangle,
\end{equation}
where $p = \sum_{l_b=0}^N\sum_{l_a=0}^{N-l_b}\sum_{n=l_b}^{N-l_a}\left|A_n\right|^2B_{l_a,l_b}^n$ is a normalization constant for state $\left|\Psi_{out}\right\rangle$.

\begin{figure}[t]
\center{\includegraphics[width=0.5\linewidth]{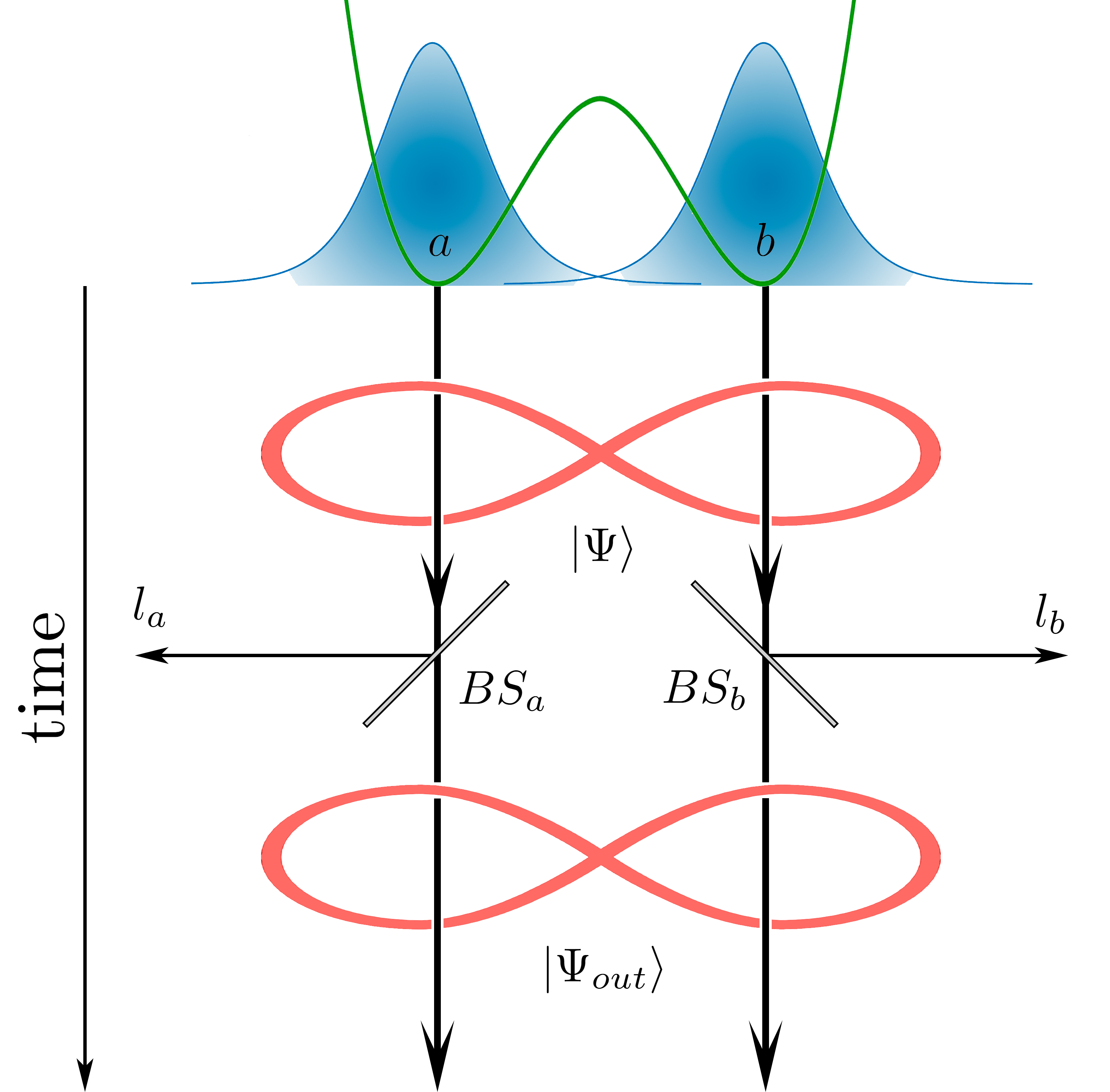}}
\caption{Scheme of particle losses from the condensate solitons trapped in double-well potential. Interaction with environment at each well are provided by means of fictitious beam splitters $BS_a$ and $BS_b$ in quantum channels ``a" and ``b", respectively.}
\label{FIG:Particle_losses}
\end{figure}

Let us suppose that we can somehow efficiently detect the numbers, $l_{a,b}$, of particles leaving the condensate state. In particular, this situation is similar to the conditional state preparation by means of subtraction of a small number of particles. For example, in quantum optical experiments real BS devices are used instead of fictitious ones, cf.~\cite{Ourjoumtsev2006,Vitelli2010}. Consider $l_a=1$ and $l_b=0$, i.e. if only one particle is detected in channel ``a" after $BS_a$, the conditional state $|\Psi^{(1,0)}_{out}\rangle$ obtained from~\eqref{out} at $\eta_a=\eta_b=\eta$ takes the form
\begin{equation}\label{out01}
\left|\Psi^{(1,0)}_{out}\right\rangle = \frac{1}{\sqrt{p}}\sum_{n=0}^{N-1}A_n\sqrt{N-n}\sqrt{\eta^{N-1}(1-\eta)}\left|N-n-1\right\rangle_a\left|n\right\rangle_b.
\end{equation}

In Fig. 7 we represent the probabilities $|C^n_{1,0}|^2$ ( $C^n_{l_a,l_b}\equiv \frac{1}{\sqrt{p}}A_n\sqrt{B_{l_a,l_b}^n}$) for the SJJ- and BJJ-models versus $n$ in the presence of weak and equal losses in each channel, $\eta_a=\eta_b=0.999$. To be more specific, in Fig.7(a,b), we use the same condensate parameters as in Fig.3(c,d), respectively.

The multiplier $\sqrt{N-n}$ in~\eqref{out01} plays a crucial role in the behavior of relevant probability $|C^n_{1,0}|^2$ that characterizes the conditional state $|\Psi^{(1,0)}_{out}\rangle$. In the presence of weak losses, the tendency to occupation of ``edge" states continues with increasing of parameter $\Lambda$, see Fig.7(a). However, in this case the state with $n=N$ cannot be occupied due to one particle loss, cf. Fig.~\ref{FIG:ground_states}(c). Strictly speaking, the state with $n=0$, which is state $|N-1\rangle_a|0\rangle_b$, becomes macroscopically populated for the SJJ-model with probability $|C^n_{1,0}|^2\propto \sqrt{N}\left|A_{N-1}\right|^2\eta^{N-1}(1-\eta)$. On the contrary, the probability of occupation of new ``edge" state with $n=N-1$ is more than $\sqrt{N}$ times lower. Physically, this case may by recognized as so-called partial ``collapse" of $N00N$-state~\eqref{N00N2} to state $|N-1\rangle_a|0\rangle_b$, when one particle is detected in ``a" channel; some small occupation of the state with $n=N-1$ still exists, see the inset in Fig. 7(b). Notice that the states with $n=N-1$, $N-2,...$ are poorly populated even without any losses - see insets to Fig. 3(c) and Fig. 3(d).

It is worth noticing that since the particles possess some (Gaussian-like) distribution in the vicinity of ``edges", the SC-states for the BJJ-model seem to be more robust to the particle losses, see the inset in Fig. 7(d) and cf. Fig. 3(d).

However, for $l_a=l_b=1$, i.e., if two particles are detected simultaneously after BSs in Fig. 6, the behavior of the conditional state (55) becomes fundamentally different. In this case for (55) we obtain:
\begin{equation}\label{out11}
\left|\Psi^{(1,1)}_{out}\right\rangle = \frac{1}{\sqrt{p}}\sum_{n=1}^{N-1}A_n \sqrt{n(N-n)\left(\frac{\eta_b}{\eta_a}\right)^n\eta_a^N(\eta_a^{-1}-1)(\eta_b^{-1}-1)}\left|N-n-1\right\rangle_a\left|n-1\right\rangle_b.
\end{equation}
In this limit, new ``edge" states with $n=1$ and $n=N-1$ are occupied with comparable probabilities. In particular, for $\eta_a\neq\eta_b$ superposition (57) turns to the unbalanced $N00N$-state that possesses $N-2$ total number of particles and has a form
\begin{equation}\label{N00n11}
\left|\Psi^{(1,1)}_{N00N}\right\rangle \propto \sqrt{(N-1)(1-\eta_a)(1-\eta_b)} \left(\sqrt{\eta_a^{N-2}}\left|N-2\right\rangle_a\left|0\right\rangle_b +\sqrt{\eta_b^{N-2}}\left|0\right\rangle_a\left|N-2\right\rangle_b\right).
\end{equation}

The unbalanced $N00N$-state (58) represents a useful tool for the quantum metrology purposes in the presence of weak losses if the corresponding transmissivities exceed a certain threshold value, cf.~\cite{Dobrzanski2009}.

To generalize these results, let us examine the state (55) in the case, when the numbers of lost particles, $l_a$ and $l_b$, are undetected (and unknown), as it is shown in Fig. 6. Tracing out the ancillary modes $l_a$, $l_b$ in (55) for the resulting state we obtain (cf.~\cite{Vogel2015})
\begin{equation}\label{out_cond}
\left|\Psi_{out}\right\rangle = \sum_{l_b=0}^N\sum_{l_a=0}^{N-l_b}\sum_{n=l_b}^{N-l_a}C_{l_a,l_b}^n\left|N-n-l_a\right\rangle_a\left|n-l_b\right\rangle_b.
\end{equation}
In Fig.~\ref{FIG:3d_Bars}, we plot the probabilities $|C^n_{l_a,l_b}|^2$ in $N-n-l_a$ and $n-l_b$ axis for the SJJ-model. Each column in Fig.~\ref{FIG:3d_Bars} corresponds to the probability of state $\left|N-n-l_a\right\rangle_a\left|n-l_b\right\rangle_b$ occupation. The parameter $\Lambda=4$ in Fig.~\ref{FIG:3d_Bars} is the same as for Fig.~\ref{FIG:ground_states}(d). Two big (yellow) columns in Fig.~\ref{FIG:3d_Bars} correspond to the $N00N$-state if no condensate particles are lost. In general, in the absence of losses the occupied modes are located on the main diagonal between these columns. In the presence of losses, the terms with $l_a\neq0$ and/or $l_b\neq0$ appear; they contain $N' = N-l_a-l_b$ particles and are characterized by the columns on the lines parallel to the main diagonal in Fig.~\ref{FIG:3d_Bars}. In particular, the columns depicted in Orange and Green-colors in Fig.~\ref{FIG:3d_Bars} form the set of the $N00N$-states with $N'<N$ number of particles, see (58) and cf.~\cite{Vogel2015}. These superposition states still exist if equal numbers of particle $l_a$ and $l_b$ are measured after BSs in Fig. 6. However, relevant probabilities essentially vanish with increasing $l_a$ and $l_b$.

\begin{figure*}[t]
\center{\includegraphics[width=1\linewidth]{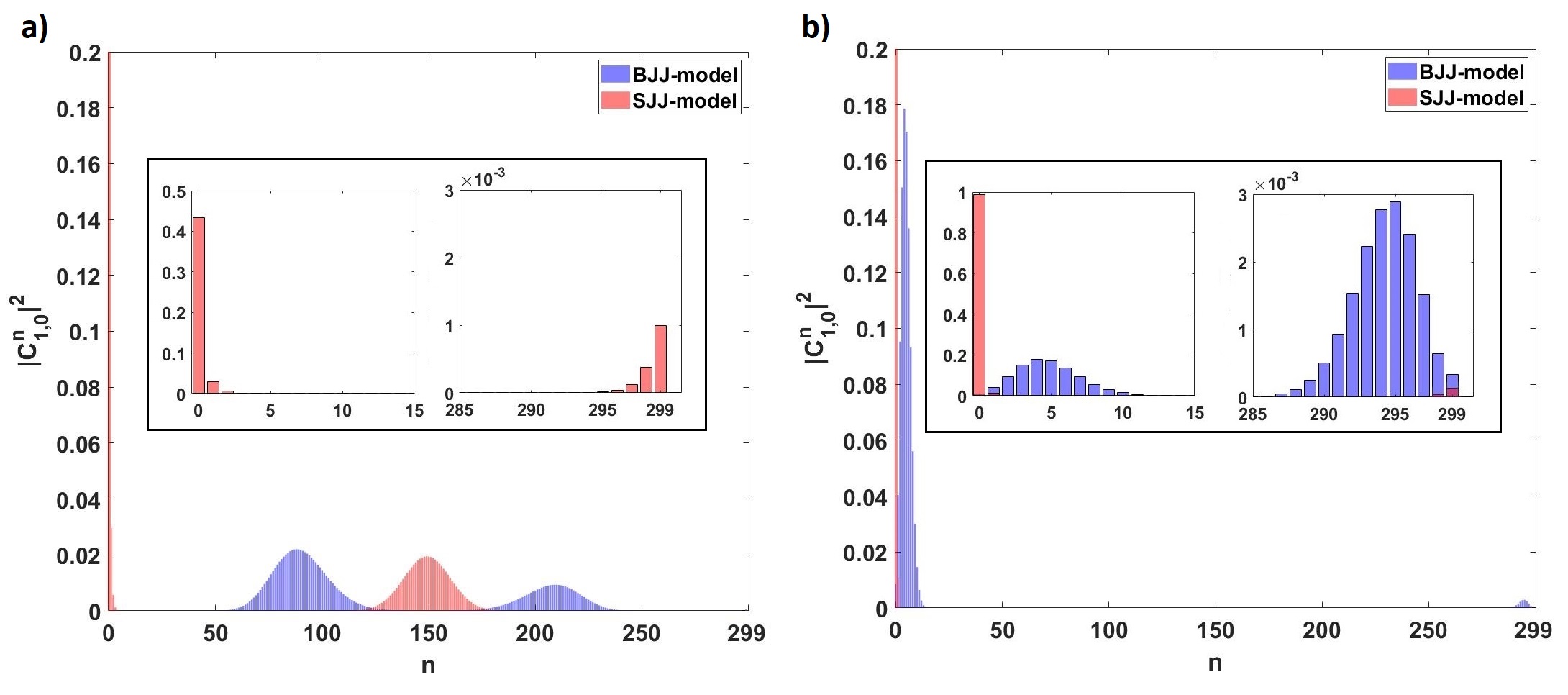}}
\caption{Probabilities $|C^n_{1,0}|^2$ vs. $n$ for (a) $\Lambda\approx 2.0009925$, $\lambda=1.06$; (b) $\Lambda=\lambda = 4$. The parameters are: $\eta_a=\eta_b=0.999$, $l_a=1$, $l_b=0$. Initial total particle number is $N=300$. Insets show probability features in the vicinity of ``edge" states with $n=0$ (left-hand panels) and $N'=N-1=299$ (right-hand panels), respectively.}
\label{FIG:2d_Bars_losses}
\end{figure*}

\begin{figure}[t]
\center{\includegraphics[width=0.75\linewidth]{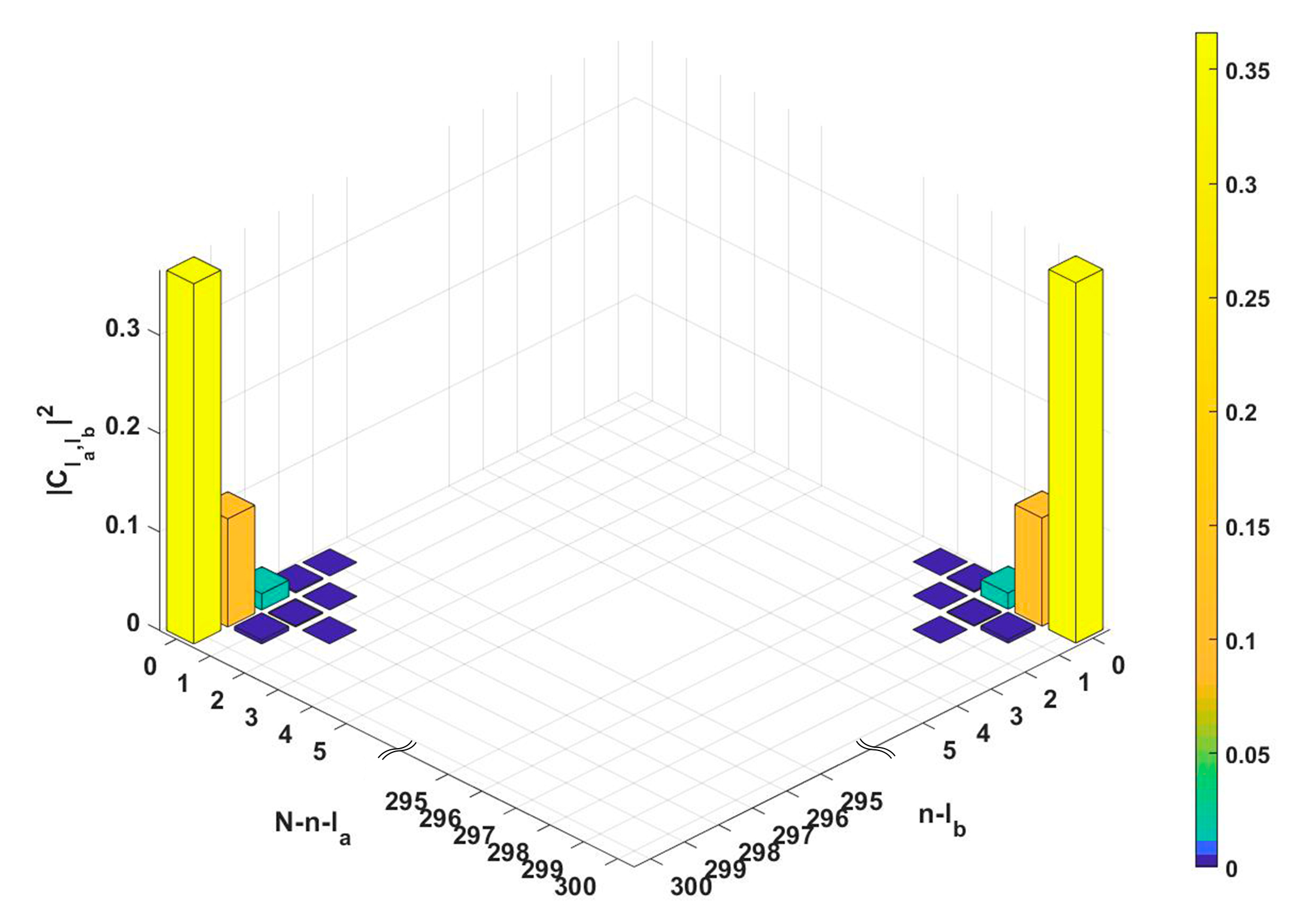}}
\caption{The 3D probabilities $|C^n_{l_a,l_b}|^2$ versus $N-n-l_a$ and $n-l_b$ for $\eta_a = \eta_b = 0.999$ and $\Lambda = 4$, $N=300$.}
\label{FIG:3d_Bars}
\end{figure}

\section{Conclusion}

In this work we have considered the problem of macroscopic and mesoscopic states formation with two weakly-coupled atomic condensate bright solitons. The condensates are trapped in a double-well potential in the $YZ$-plane and are significantly elongated in the $X$-axis forming the specific Josephson junctions (JJs) geometry. We assume that condensates admit a weak interaction between the atoms possessing a negative scattering length of atom-atom interaction. This geometry allows to examine bright atomic soliton features in one ($X$) dimension. We start from the classical (mean) field approach to the problem that admits bright soliton solutions with preserving soliton shapes. In this limit these solutions permit to effectively map the system of weakly-coupled solitons onto the two-mode problem. 

Transferring to the quantum picture of the problem we perform the second quantization procedure of effectively interacting modes that allows us to obtain the quantum version of the classical two-mode Hamiltonian. We refer such a Hamiltonian to the soliton Josephson junction (SJJ) model.

Comparing all the results obtained for the SJJs with the ones of Bosonic Josephson junction (BJJ) model consisting of two weakly-coupled (Gaussian-shape) condensates, we have shown that there exist two important differences between the SJJ- and BJJ-models. First, the SJJ-model permits a particle dependent tunneling rate. Second, the effective nonlinear parameter for the SJJ-model depends on the total particle number squared, $N^2$, not $N$, that takes place for the BJJ-model. To demonstrate the advantages of the SJJ-device we have examined the quantum analysis of the superposition states formation. 

Initially, we have considered the quantum field variational (Hartree) approach that predicts Schr{\"o}dinger-cat states formation within the vital parameter, $\Lambda$, domain. In this limit we explore the atomic coherent state as the basis for the superposition state analysis. We have shown that the cat states approach the $N00N$-state in the limit when the cat becomes macroscopic. 

We have performed the full quantum analysis of the energy spectrum for the SJJ- and BJJ-models to consider the $N00N$-state formation problem for the JJs in detail. We have demonstrated that the crossover to the $N00N$-state may be realized for the both models at some combination of atomic nonlinear strength and tunneling rate. However, due to the particle dependence, the effective tunneling rate, $\kappa_{eff}$, that governs this crossover, demonstrates self-tuning effect, revealing an abrupt transition to superposition of entangled Fock states. The $N00N$-state appears at the ``edges" with $n=0,N$ of this superposition and represents its major component. The existence of the satellite entangled Fock state components with non-zero probabilities in the vicinity of ``edges" represent an important difference from the variational (Hartree) approach. It is worth noting that probabilities of these components are vanishing, and the SJJ-model approaches approximately ``ideal" $N00N$-state (45) for moderate values of $\Lambda$-parameter. The BJJ-model approaches the same $N00N$-state only with a quite large nonlinear strength, $\lambda\geq60$. 

To confirm the results obtained we have examined the $m$-order Hillery-Zubairy criterion and taken it with $m=1$ and $m=N$, respectively. The numerical calculations reflect the existence of the strong entanglement and the planar quantum squeezing, which occur up to the crossing point defined by the vital parameter $\Lambda$ (or $\lambda$). The EPR steering happens in the crossover point vicinity where Hartree approach is useless. Notably, the depth of the steering is much higher for the familiar atomic JJs, which may be explained through the additional particle number fluctuations occurring from the tunneling process for the SJJ-model. This process leads to a narrower distribution function appearing immediately at the crossover point. The results obtained for the HZ-criterion indicate the crossover from the quantum steering to the $N00N$-state. This crossover exhibits a very sharp and large variation in the first order HZ-criterion that takes place for the SJJ-model. 

Finally, we examine how losses influence the formation of state (37). The qualitative estimations of influence from one- and three-body losses demonstrate feasibility of observation on predicted phenomena with SJJs. To take into account the losses of a few number of particles in condensate solitons, we have used the fictitious beam splitter approach that represents the essence of the state preparation procedure in Fig. 6. In particular, we have shown that the existence of tiny satellite entangled Fock state components provides some resistance of the $N00N$-state to complete collapsing if we are able to detect (non-equal) number particles which are lost in each well (channel). We have also shown that the final state (56) preserves the $N00N$-state structure if two or even number of particles are lost and may be equally detected in both channels. Especially, such an analysis seems to be important for practical applications of the soliton JJs in ``real world" quantum metrology.

We are confident that the quantum state features of the SJJ-model described by Hamiltonian in Eq. (20) potentially may be significantly larger than we describe in this work. In particular, established  quantum SJJ-model may be also useful in quantum communication and quantum information science, where quantum optical solitons represent an appropriate tool, cf.~\cite{Haus2000}. In this sense quantum properties of the SJJ-model proposed in this work represent a necessary step to further studies.

This work was financially supported by the Grant of RFBR, No. 19-52-52012 $MHT-a$. RKL was supported by the Ministry of Science and Technology of Taiwan (No. 108-2923-M-007-001-MY3 and
No. 109-2112-M-007-019-MY3). AA acknowledges partial support from Leading Research Center "National Center of Quantum Internet" of ITMO University in the framework of Grant Agreement ID:
0000000007119P190002, agreement No. 006-20 dated 27.03.2020.


\section*{References}
\begin{thebibliography}{58}
\bibitem{Josephson1974} 
Josephson B D 1974 \textit{Review of Modern Physics} \textbf{46} 251–-254

\bibitem{Pereverzev1997} 
Pereverzev S V, Loshak A, Backhaus S, Davis J C, and Packard R E 1997 \textit{Nature} \textbf{388} 449

\bibitem{Davis2002} 
Davis J C and Packard R E 2002 \textit{Review of Modern Physics} \textbf{74} 741

\bibitem{Sukhatme2001} 
Sukhatme K, Mukharsky Y, Chui T, and Pearson D 2001 \textit{Nature} \textbf{411} 280

\bibitem{Leggett2001}
Leggett A J 2001 \textit{Review of Modern Physics} \textbf{73} 307

\bibitem{Albiez2005}
Albiez M, Gati R, F{\"o}lling J, Hunsmann S, Cristiani M, and Oberthaler M K 2005 \textit{Physical Review Letters} \textbf{95} 010402

\bibitem{Milburn1997}
Milburn G J, Corney J, Wright E M, and Walls D F 1997 \textit{Physical Review A} \textbf{55} 4318

\bibitem{Raghavan1999}
Raghavan S, Smerzi A, Fantoni S, and Shenoy S R 1999 \textit{Physical Review A} \textbf{59} 620

\bibitem{Abbarchi2013}
Abbarchi M, Amo A, Sala V G, Solnyshkov D D, Flayac H, Ferrier L, Sagnes I, Galopin E, Lema{\^i}tre A, Malpuech G, and Bloch J 2013 \textit{Nature Physics} \textbf{9} 275--279

\bibitem{Lagoudakis2010}
Lagoudakis K G, Pietka B, Wouters M, Andr{\'e} R, and Deveaud-Pl{\'e}dran B 2010 \textit{Physical Review Letters} \textbf{105} 120403

\bibitem{Lebedev2017}
Lebedev M E, Dolinina D A, Hong K-B, Lu T-C, Kavokin A V, and Alodjants A P 2017 \textit{Scientific Reports} \textbf{7} 1--12

\bibitem{Maier1995}
Maier A A 1995 \textit{Physics-Uspekhi} \textbf{38} 991–-1029

\bibitem{Leksin2003}
Leksin A Y, Alodjants A P, and Arakelian S M 2003 \textit{Journal of Russian Laser Research} \textbf{24} 168--179

\bibitem{Cirac1998}
Cirac J I, Lewenstein M, M{\o}lmer K, and Zoller P 1998 \textit{Physical Review A} \textbf{57} 1208

\bibitem{Huang2006}
Huang Y P and Moore M G 2006 \textit{Phys. Rev. A} \textbf{73} 023606

\bibitem{Olsen2010}
Haigh T J, Ferris A J, and Olsen M K 2010 \textit{Optics Communications} \textbf{283} 3540--3547

\bibitem{Reid2011}
He Q Y, Peng S G, Drummond P D, and Reid M D 2011 \textit{Phys. Rev. A} \textbf{84} 022107

\bibitem{Salasnich2011}
Mazzarella G, Salasnich L, Parola A, and Toigo F 2011 \textit{Phys. Rev. A} \textbf{83} 053607

\bibitem{Sorensen2001}
S{\o}rensen A S and M{\o}lmer K 2001 \textit{Phys. Rev. Letts.} \textbf{86} 4431

\bibitem{Sorensen2001a} 
S{\o}rensen A S, Duan L-M, Cirac J I, and Zoller P 2001 \textit{Nature} \textbf{409} 63--66

\bibitem{Reid2012}
He Q Y, Drummond P D, Olsen M K, and Reid M D 2012 \textit{Phys. Rev. A} \textbf{86} 023626

\bibitem{Vitagliano2018}
Vitagliano G, Colangelo G, MartinCiurana F, Mitchell M W, Sewell R J, and Toth G 2018 \textit{Phys. Rev. A} \textbf{97} 020301(R)

\bibitem{Reid2018}
Rosales-Z{\'a}rate L, Dalton B J, and Reid M D 2018 \textit{Phys. Rev. A} \textbf{98} 022120

\bibitem{Maa2011}
Maa J, Wang X, Sun C P, and Nori F 2011 \textit{Physics Reports} \textbf{509} 89--165

\bibitem{Jones2007} 
Jones S J, Wiseman H M, and Doherty A C 2007 \textit{Phys. Rev. A} \textbf{76} 052116

\bibitem{Leggett1984}
Leggett A J 1984 \textit{Contemp. Phys.} \textbf{25} 583--598

\bibitem{Castanos2012}
Casta{\~n}os O, and L{\'o}pez-Sald{\'i}var J A 2012 \textit{J. of Phys.: Conf. Ser.} \textbf{380} 012017

\bibitem{Castanos2020}
L{\'o}pez-Sald{\'i}var J A 2020 \textit{Phys. Scr.} \textbf{95} 065206

\bibitem{Byrnes2012}
Byrnes T, Wen K, and Yamamoto Y 2012 \textit{Physical Review A} \textbf{85} 040306(R)

\bibitem{Byrnes2011} 
Byrnes T, Yan K, and Yamamoto Y 2011 \textit{New J. of Phys.} \textbf{13} 113025

\bibitem{Pezze2018} 
Pezz{\`e} L, Smerzi A, Oberthaler M K, Schmied R, and Treutlein P 2018 \textit{Rev. of Modern Phys.} \textbf{90} 035005

\bibitem{Puentes2013} 
Puentes G, Colangelo G, Sewell R J, and Mitchell M W 2013 \textit{New J. of Phys.} \textbf{15} 103031 

\bibitem{Dowling2008} 
Dowling J P 2008 \textit{Contemporary Physics} \textbf{49} 125–-143

\bibitem{Afek2010}
Afek I, Ambar O, and Silberberg Y 2010 \textit{Science} \textbf{328} 879–-881

\bibitem{Chen2010}
Chen Y-A, Bao X-H, Yuan Z-S, Chen S, Zhao B, and Pan J W 2010 \textit{Phys. Rev. Letts.} \textbf{104} 043601

\bibitem{Merkel2010}
Merkel S T and Wilhelm F K 2010 \textit{New J. of Phys.} \textbf{12} 093036

\bibitem{Dorner2009}
Dorner U, Demkowicz-Dobrzanski R, Smith B J, Lundeen J S, Wasilewski W, Banaszek K, and Walmsley I A 2009 \textit{Phys. Rev. Letts.} \textbf{102} 040403

\bibitem{Vogel2015}
Bohmann M, Sperling J, and Vogel W 2015 \textit{Phys. Rev. A} \textbf{91} 042332

\bibitem{Huver2008}
Huver S D, Wildfeuer C F, and Dowling J P 2008 \textit{Phys. Rev. A} \textbf{78} 063828

\bibitem{Dobrzanski2009}
Demkowicz-Dobrzanski R, Dorner U, Smith B J, Lundeen J S, Wasilewski W, Banaszek K, and Walmsley I A 2009 \textit{Phys. Rev. A} \textbf{80} 013825

\bibitem{Sinatra2017}
Paw{\l}owski K, Fadel M, Treutlein P, Castin Y, and Sinatra A 2017 \textit{Phys. Rev. A} \textbf{95} 063609

\bibitem{Tsarev2018}
Tsarev D V, Arakelyan S M, Chuang Y-L, Lee R-K, and Alodjants A P 2018 \textit{Optics Express} \textbf{26} 19583 

\bibitem{Tsarev2019}
Tsarev D V, Ngo T.V., Lee R-K, and Alodjants A P 2019 \textit{New J. of Phys.} \textbf{21} 083041

\bibitem{Raghavan2000}
Raghavan S and Agrawal G P 2000 \textit{Journal of Modern Optics} \textbf{47} 1155--1169

\bibitem{Lieb1963}
Lieb E H and Liniger W 1963 \textit{Phys. Rev.} \textbf{130} 1605–-1616 (1963)

\bibitem{Haus1989}
Lai Y and Haus H A 1989 \textit{Phys. Rev. A} \textbf{40} 844 

\bibitem{Zill2015}
Zill J C, Wright T M, Kheruntsyan K V, Gasenzer T, and Davis M J 2015 \textit{Phys. Rev. A} \textbf{91} 023611

\bibitem{Drummond1987}
Carter S, Drummond P, Reid M, and Shelby R 1987 \textit{Phys. Rev. Letts.} \textbf{58} 1841

\bibitem{Friberg1996}
Friberg S R, Machida S, Werner M J, Levanon A, and Mukai T 1996 \textit{Phys. Rev. Letts.} \textbf{77} 3775

\bibitem{Spalter1998}
Sp{\"a}lter S, Korolkova N, K{\"o}nig F, Sizmann A, and Leuchs G 1998 \textit{Phys. Rev. Letts.} \textbf{81} 786

\bibitem{Lai2009}
Lai Y and Lee R-K 2009 \textit{Phys. Rev. Letts.} \textbf{103} 013902

\bibitem{Kevrekidis2008}
Kevrekidis P G, Frantzeskakis D J, Carretero-Gonz{\'a}lez R 2008 \textit{Emergent Nonlinear Phenomena in Bose–Einstein Condensates} (Springer-Verlag Berlin Heidelberg)

\bibitem{Strecker2002}
Strecker K E, Partridge G B, Truscott A G, and Hulet R G 2002 \textit{Nature} \textbf{417} 150-–153

\bibitem{Khaykovich2002}
Khaykovich L, Schreck F, Ferrari G, Bourdel T, Cubizolles J, Carr L D, Castin Y, and Salomon C 2002 \textit{Science} \textbf{296} 1290-–1293

\bibitem{Nguyen2014}
Nguyen J, Dyke P, Luo D, Paul Dyke, De Luo, Malomed B A, and Hulet R G 2014 \textit{Nature Phys.} \textbf{10} 918-–922

\bibitem{Eiermann2004}
Eiermann B, Anker Th. Albiez M, Taglieber M, Treutlein P, Marzlin K-P, and Oberthaler M K 2004 \textit{Phys. Rev. Letts.} \textbf{92} 230401

\bibitem{Weiss2016}
Weiss Ch, Cornish S L, Gardiner S A, and Breuer H P 2016 \textit{Phys. Rev. A} \textbf{93} 013605

\bibitem{Carr2004}
Carr L D, Brand J 2004 \textit{Phys. Rev. A} \textbf{70} 033607

\bibitem{Malomed2006}
Malomed B A 2006 \textit{Soliton management in periodic systems} (USA: Springer Science \& Business Media)

\bibitem{Gardiner2012}
Holdaway D I H, Weiss C, and Gardiner S A 2012 \textit{Phys. Rev. A} \textbf{85} 053618

\bibitem{Cabrera2018}
Cabrera C R, Tanzi L, Sanz J, Naylor B, Thomas P, Cheiney P, and Tarruell L 2018 \textit{Science} \textbf{359} 301–-304

\bibitem{Pfau2016}
Ferrier-Barbut I, Kadau H, Schmitt M, Wenzel M, and Pfau T. 2016 \textit{Phys. Rev. Letts.} \textbf{116} 215301

\bibitem{Petrov2016}
Petrov D S and Astrakharchik G E 2016 \textit{Phys. Rev. Letts.} \textbf{117} 100401

\bibitem{Liu2019}
Liu B, Zhang H-F, Zhong R X, Zhang X-L, Qin X Z, Huang Ch, Li Y-Y, and Malomed B A 2019 \textit{Phys. Rev. A} \textbf{99} 053602

\bibitem{Tylutki2020}
Tylutki M, Astrakharchik G E, Malomed B A, and Petrov D S, 2020 \textit{Phys. Rev. A} \textbf{101} 051601(R)

\bibitem{Rubin2007}
Rubin M A, and Kaushik S 2007 \textit{Phys. Rev. A} \textbf{75} 053805

\bibitem{Gati2007}
Gati R and Oberthaler M K, 2007 \textit{J. Phys. B: At. Mol. Opt. Phys.} \textbf{40} R61

\bibitem{Anglin2001}
Anglin J R and Vardi A 2001 \textit{Physical Review A} \textbf{64} 013605

\bibitem{Elef2000}
Eleftheriou A and Huang K 2000 \textit{Phys. Rev. A} \textbf{61} 043601

\bibitem{Eigen2016}
Eigen Ch, Gaunt A L, Suleymanzade A, Navon N, Hadzibabic Z, and Smith R P 2016 \textit{Phys. Rev. X} \textbf{6} 041058

\bibitem{Pearl2003}
Louis P J, Ostrovskaya E A, Savage C M, and Kivshar Y S, 2003 \textit{Phys. Rev. A} \textbf{67}, 013602

\bibitem{Cheng2010}
Chin C, Grimm R, Julienne P, and Tiesinga E 2010 \textit{Review of Modern Physics} \textbf{82} 1225

\bibitem{Sedov2014}
Charukhchyan M V, Sedov E S, Arakelian S M, and Alodjants A P 2014 \textit{Phys. Rev. A} \textbf{89} 063624

\bibitem{Skott1992}
Enol'skii V Z, Salerno M, Scott A C, and Eilbeck J C 1992 \textit{Physica D: Nonlinear Phenomena} \textbf{59} 1--24 

\bibitem{Brand2010}
Brand J, Haigh T J, and Z{\"u}licke U 2010 \textit{Phys. Rev. A} \textbf{81} 025602

\bibitem{Leggett2011}
Paraoanu G-S, Kohler S, Sols F, and Leggett A J 2001 \textit{Journal of Physics B} \textbf{34} 4689–-4696

\bibitem{Alodjants1995} 
Alodjants A P and Arakelian S M 1995 \textit{JETP} \textbf{80} 995--1012

\bibitem{Wright1993}
Wright E, Eilbeck J C, Hays M H, Miller P D, and Scott A C 1993 \textit{Physica D: Nonlinear Phenomena} \textbf{69} 18--32 

\bibitem{Colzi2018}
Colzi G, Fava E, Barbiero M, Mordini C, Lampores G, and Ferrari G 2018 \textit{Phys. Rev. A} \textbf{97} 053625

\bibitem{Lewenstein2007}
Lewenstein M, Sanpera A, Ahufinger V, Damski B, Sen A, and Sen U 2007 \textit{Adv. in Phys.} \textbf{56} 243-–379

\bibitem{Arndt2009}
Greenberger D, Hentschel K, and Weinert F 2009 \textit{Compendium of Quantum Physics} (Springer-Verlag Berlin Heidelberg)

\bibitem{Kapale2007}
Kapale K T, Didomenico L D, Lee H, Kok P, and Dowling J P 2007 \textit{Proc. of SPIE} \textbf{6603} 660316

\bibitem{Knoop2012}
Knoop S, Borbely J S, van Rooij R, and Vassen W 2012 \textit{Phys. Rev. A} \textbf{85} 025602

\bibitem{Li2008}
Li Y, Castin Y, and Sinatra A 2008 \textit{Phys. Rev. Letts.} \textbf{100} 210401

\bibitem{Greiner2002}
Greiner M, Mandel O, Esslinger T, H{\:a}nsch T, and Bloch I 2002 \textit{Nature} \textbf{415} 39--44

\bibitem{Andreev1998}
Andreev A F 1998 \textit{Physics-Uspekhi} \textbf{41} 581

\bibitem{Messiah1968} 
Messiah A 1968 \textit{Quantum Mechanics: volume 1} (NY: John Wiley and Sons).

\bibitem{Fedichev1996}
Fedichev P O, Reynolds M V, and Shlyapnikov G V 1996 \textit{Phys. Rev. Letts.} \textbf{77} 2921

\bibitem{Grimm2003}
Weber T, Herbig J, Mark M, N{\:a}gerl H C, and Grimm R 2003 \textit{Phys. Rev. Letts.} \textbf{91} 123201

\bibitem{Sols2002}
Kohler S and Sols F 2002 \textit{Phys. Rev. Letts.} \textbf{89} 060403

\bibitem{Caldeira1983}
Caldeira A O and Leggett A J 1983 \textit{Ann. of Phys.} \textbf{149} 374--456

\bibitem{Verhaar1996}
Moerdijk A J, Boesten H M J M, Verhaar B J 1996 \textit{Phys. Rev. A} \textbf{53} 916

\bibitem{Gross2009}
Gross N, Shotan Z, Kokkelmans S, and Khaykovich L 2009 \textit{Phys. Rev. Letts.} \textbf{103} 163202

\bibitem{Petrov2013}
Rem B S, Grier A T, Ferrier-Barbut I, Eismann U, Langen T, Navon N, et. al. 2013 \textit{Phys. Rev. Letts.} \textbf{110} 163202

\bibitem{Ourjoumtsev2006}
Ourjoumtsev A, Tualle-Brouri R, Laurat J, and Grangier P 2006 \textit{Science} \textbf{312} 83--86

\bibitem{Vitelli2010}
Vitelli C, Spagnolo N, Toffoli L, Sciarrino F, and De Martini F 2010 \textit{Phys. Rev. Letts.} \textbf{105} 113602

\bibitem{Haus2000}
Haus H A \textit{Electromagnetic Noise and Quantum Optical Measurements} (Springer-Verlag Berlin Heidelberg) 501

\end{thebibliography}
\end{document}